\documentclass{interact}

\usepackage{epstopdf}
\usepackage{subfigure}
\usepackage{cancel}
\usepackage{hyperref}

\usepackage[numbers,sort&compress]{natbib}
\bibpunct[, ]{[}{]}{,}{n}{,}{,}
\makeatletter
\def\NAT@def@citea{\def\@citea{\NAT@separator}}
\makeatother

\usepackage[T1]{fontenc} 
\usepackage{lipsum}
\usepackage{cancel}
\usepackage{graphicx}
\usepackage{subcaption} 
\usepackage{hyperref}
\usepackage{scalerel}
\usepackage{tikz}
\usepackage{soul}

\def\vep{\varepsilon}

\def\vx{\vec{x}}
\def\vr{\vec{r}}
\def\vE{\vec{E}}
\def\z{\zeta}
\def\vk{\vec{k}}
\def\vka{\vec{\kappa}}
\def\vp{\vec{p}}
\def\vq{\vec{q}}
\def\vap{\varphi}
\def\mJ{\mathcal{J}}
\def\Upa{\uparrow}
\def\Dwa{\downarrow}

\def\gx{\gamma_{,x}}
\def\gy{\gamma_{,y}}
\def\gz{\gamma_{,\z}}

\def\nn{\nonumber\\}
\def\beq{\begin{equation}}
\def\teq{\end{equation}}

\usetikzlibrary{svg.path}
\def\BibTeX{{\rm B\kern-.05em{\sc i\kern-.025em b}\kern-.08em
    T\kern-.1667em\lower.7ex\hbox{E}\kern-.125emX}}
\usepackage{balance}

\definecolor{orcidlogocol}{HTML}{A6CE39}
\tikzset{
  orcidlogo/.pic={
    \fill[orcidlogocol] svg{M256,128c0,70.7-57.3,128-128,128C57.3,256,0,198.7,0,128C0,57.3,57.3,0,128,0C198.7,0,256,57.3,256,128z};
    \fill[white] svg{M86.3,186.2H70.9V79.1h15.4v48.4V186.2z}
                 svg{M108.9,79.1h41.6c39.6,0,57,28.3,57,53.6c0,27.5-21.5,53.6-56.8,53.6h-41.8V79.1z M124.3,172.4h24.5c34.9,0,42.9-26.5,42.9-39.7c0-21.5-13.7-39.7-43.7-39.7h-23.7V172.4z}
                 svg{M88.7,56.8c0,5.5-4.5,10.1-10.1,10.1c-5.6,0-10.1-4.6-10.1-10.1c0-5.6,4.5-10.1,10.1-10.1C84.2,46.7,88.7,51.3,88.7,56.8z};
  }
}

\newcommand\orcidicon[1]{\href{https://orcid.org/#1}{\mbox{\scalerel*{
\begin{tikzpicture}[yscale=-1,transform shape]
\pic{orcidlogo};
\end{tikzpicture}
}{|}}}}

\title{Non-additive surface and volume backscattering from a rough interface over a correlated random medium}

\begin{document}
\articletype{ARTICLE}
\author{
\name{Mariano Franco \orcidicon{0000-0001-7611-1688}\,\textsuperscript{a}\thanks{CONTACT Mariano Franco. Email: mfranco@iafe.uba.ar} and Esteban Calzetta \orcidicon{0000-0002-3083-3420}\textsuperscript{b,c}}
\affil{\textsuperscript{a}Instituto de Astronom\'ia y F\'isica del Espacio (UBA-CONICET), Buenos Aires, Argentina;
\textsuperscript{b}Universidad de Buenos Aires, Facultad de Ciencias Exactas y Naturales, Departamento de F\'isica, Buenos Aires, Argentina;
\textsuperscript{c}CONICET-Universidad de Buenos Aires, Instituto de F\'isica de Buenos Aires (IFIBA), Buenos Aires, Argentina}
}

\maketitle

\begin{abstract}
We compute the first-order incoherent backscattering cross-section of a rough surface
over a half-space with randomly fluctuating permittivity, when the two random fields are
correlated, so that surface and volume scattering are not additive. A non-orthogonal
coordinate transformation flattens the rough boundary and turns the geometric roughness
into volume sources of the same Helmholtz equation as the dielectric fluctuations,
propagated by the same Green's function: both mechanisms and their interference follow
from one calculation, with no equivalent surface currents and no prescription for the
field discontinuity. The geometric part of the 
amplitude reproduces the small-perturbation kernel exactly, for any incidence angle,
permittivity and profile function. The
interference term is obtained for both co-polarized channels, with an explicit
surface-dielectric correlation coefficient. The departure from additivity reaches twenty
percent at full correlation, changes sign with that coefficient, and is governed by a
balance condition and a coherence condition on the depth of the correlated layer. It
carries no angular signature and shifts the co-polarized ratio by less than a tenth of a
decibel, so an additive inversion absorbs it, biasing the retrieved amplitude of the
dielectric fluctuations by tens of percent.
\end{abstract}

\begin{keywords}
Rough surface scattering; random media; volume scattering; small perturbation method;
non-orthogonal coordinates; radar backscattering
\end{keywords}

\section{Introduction} \label{sec:intro}

The study of electromagnetic scattering from random rough surfaces and inhomogeneous media is a fundamental problem in classical electrodynamics with wide-ranging applications in remote sensing, geophysics, and environmental monitoring. In natural environments, such as soil layers or snow packs, the scattering process is governed by the complex interplay between the irregular topography of the interface and the stochastic fluctuations of the dielectric properties of the underlying medium. Accurate modeling of these phenomena is crucial for the inversion of geophysical parameters from radar and radiometric data, where the interface geometry and material inhomogeneities often dominate the signal.

In most treatments, these scattering mechanisms are handled as independent processes. The Small Perturbation Method (SPM) and the Kirchhoff Approximation (KA) are well-established frameworks for modeling surface scattering from slightly and moderately rough interfaces, respectively \cite{tsang2004scattering,johnson1999third,ulaby1986microwave,beckmann1987scattering,thorsos1988validity,elfouhaily2004critical,voronovich2013wave}, while volume scattering theories often assume a flat interface to simplify propagation into the medium. Under these assumptions, the total scattering cross-section is a purely additive combination of surface and volume contributions: at single scattering, the two mechanisms are incoherent, so that separate methods may be used for each and the results simply added \cite{mudaliar1995wave}. 
However, in natural media the moisture distribution in soil or the density of snow layers is correlated with the local surface height through erosion or accumulation, so the two random fields are not statistically independent. 

In a paper that remains the reference on the subject, Elson \cite{elson1984theory} applied first-order perturbation theory to a plane-bounded semi-infinite medium having both a rough boundary and a randomly fluctuating permittivity, and showed that the two scattered fields interfere, so that the scattered power depends not only on the two auto-correlations but also on the cross-correlation between surface height and dielectric fluctuations. Working in the optical regime, in \cite{elson1984theory} it is shown why the measured polarization ratio of light scattered from nominally identical silver films varies widely from sample to sample; the same problem is studied in \cite{elson1997characteristics}. A related first-order treatment of roughness and bulk-inhomogeneity scattering, built on a different formalism, was extended to anisotropic films \cite{germer2017polarized}; independently, a first-order theory of bulk scattering in multilayers was shown to take the same functional form as that of surface roughness \cite{amra1993first}.. A rough interface carrying a dielectric boundary layer had also been treated in the $T$-matrix formulation \cite{rozhnov1988electromagnetic}, and a random medium layer beneath a random interface was analyzed to first order in \cite{mudaliar1994electromagnetic}. In the microwave regime, where the target is typically a soil column of varying moisture beneath a rough interface, the problem is most often handled by combining a rough-surface model with a layered or vertically graded medium, analytically (for stratified profiles \cite{kuo2007,zamani2015scattering,imperatore2017modelling}) or numerically \cite{duan2017full,tsang2012electromagnetic}. A fully polarimetric treatment of a rough surface over an inhomogeneous medium have been developed in  \cite{pinel2011fully,yang2019full,yang2020depolarized}; general accounts of the two classes of randomness are given in \cite{ishimaru2002wave,sarabandi2002electromagnetic}. 

The main goal of this work is to derive the coupling between the two scattering mechanisms and to establish under what conditions this coupling is large enough that it cannot be disregarded.

Because first-order perturbation theory treats the two mechanisms as independent, they are customarily computed by different methods: for example, the scattering due to roughness is treated by matching boundary conditions across the interface, while the scattering from dielectric fluctuations is treated by a Born expansion with a dyadic Green's function in the bulk; later. The results are then added as a linear superposition of fields \cite{elson1984theory}. As a consequence, the interference term must be arbitrary introduced. This is the route followed both in the optical \cite{elson1984theory,elson1997characteristics} and in the microwave \cite{mudaliar1994electromagnetic} literature. 

The multiple-scattering continuation of the latter \cite{mudaliar1995wave} follows the same strategy, recasting the conditions on the rough interface as effective surface impedances on a flat reference plane; their explicit form for a given interface, however, is left for future work. A related point concerns the surface currents themselves: because they are defined exactly at the interface, where the normal field is discontinuous, their placement and strength are not fully fixed by the boundary conditions alone and \cite{elson1984theory} determines this issue with an explicit convention.

We avoid both difficulties by mapping the random boundary, defined by the rough surface $z = h(\vx)$, onto a flat interface with a non-orthogonal coordinate transformation $z=\z+h(\vx)f(\z)$. The $f$ function is chosen not only to flatten the surface but also to render the medium statistically homogeneous. We expect such a function exists because near the surface the properties of the medium depend essentially on the depth $h(\vx)-z$, while away from the surface the medium is already homogeneus in the Cartesian coordinates. A function $f(\z)$ which takes the value $1$ for $\z\approx 0$ and goes to $0$ as $\z\to -\infty$ smoothly interpolates between both regimes. To avoid a discontinuity we then continue this function to arguments $\z\ge 0$. Here the specific choice of $f$ is immaterial, we assume that it is smooth, that the relationship of $z$ to $\z$ is one-to-one, and that $f(\z)\to 0$ as $\z\to\infty$. This last condition is chosen only for convenience, it means that we need not transform the scattered fields from one set of coordinates to the other, because they agree in the asymptotic region.

This strategy was inspired by the curvilinear-coordinate (or C-) method for diffraction gratings \cite{chandezon1980new,chandezon1982multicoated,li1999rigorous}. The C-method was then generalized to arbitrary transformations \cite{plumey1999generalization}, cast in covariant form \cite{plumey1995differential}, and extended to random and aperiodic rough surfaces \cite{granet2002scattering,edee2007complex,braham2008scattering,dusseaux2008implementation}. The same idea appears in transformed-field-expansion methods \cite{nicholls2004shape}, recently applied to multilayered random media \cite{ulmer2022monte}, where a rough boundary is replaced by a flat interface plus an equivalent anisotropic medium \cite{ozgun2013transformation}. These works are numerical: the transformation yields an eigenvalue system, or a high-order recursion, to be solved exactly. Instead, we develop an analytical perturbative approach, by which we obtain closed-form first-order amplitudes. We also note that rough profiles and volume permittivity fluctuations have been treated within a single coordinate system before \cite{li1999oblique}, although for deterministic periodic media and without cross-correlation between the two. In contrast, we convert the geometric roughness into volume sources of the Helmholtz equation, of the same kind and in the same equation as the dielectric fluctuations, and propagated by the same Green's function.Within this framework, the interference between the two mechanisms is a consequence of the formulation rather than an addition to it, and the construction is not limited to first order perturbation theory.

As a general framework, rather than assuming a columnar dielectric perturbation \cite{elson1984theory}, we consider permittivity to be a three-dimensional random field with an independent vertical correlation length \cite{mudaliar1994electromagnetic,mudaliar1995wave}. Furthermore, the surface-dielectric correlation carries its own lateral and vertical scales, and the resulting cross term is worked out explicitly. It is then found that the surface-dielectric correlation is governed by a coherence condition on the depth of the correlated layer, being suppressed when that layer is either too thin to contribute or thick enough for the phase to rotate across it. This is a condition that cannot appear in a columnar model, which has no vertical scale to vary. Additionally, we keep an explicit cross-correlation coefficient $\rho_0\in[-1,1]$, which separates the strength of the coupling from the scales over which it acts, and which can be carried through to the cross-section. Keeping it explicit also locates where the non-additivity of the two mechanisms comes from. The first-order cross term is proportional to $\rho_0$, so surface and volume scattering are incoherent at single scattering when the two random fields are not correlated ($\rho_0=0$).

The formulation presented here is valid at any order in a perturbation scheme. We treat with a non-homogeneous Helmholtz differential equation for an unknown field of order $n$, where the sources are given by fields of lower order. In this work, we just keep on a first-order perturbative solution for the scattered fields. We derive the incoherent backscattering cross-section for both the TE (HH) and TM (VV) channels, verifying that the geometric part of the amplitude reproduces the small perturbation method (SPM) kernel exactly and it is independent of the explicit expression for the profile function $f(\z)$. We then give the interference term produced by the cross-correlation and quantify the error incurred by inverting (synthetic) data with an additive model.

The remainder of this manuscript is organized as follows. In Section \ref{sec:change_coord}, we introduce the coordinate transformation and the resulting changes to the metric and Helmholtz equation, and establish the interface condition. Section \ref{sec:pert_develop} describes the perturbative scheme and the zeroth-order solution for a flat interface. In Section \ref{sec:first_order_sources}, we define the first-order source terms for both polarizations and the statistical properties of the random media. Section \ref{sec:mean_values} presents the derivation of the mean scattering amplitudes and of the incoherent cross-section, including the interference term. In Section \ref{sec:results} we present the main aspect that the correlation coefficient $\rho_0$ introduces in the backscattering behavior. The conclusions are outlined in Section \ref{sec:conclusions}.

\section{Helmholtz equation: from cartesian to non-orthogonal coordinates}
\label{sec:change_coord}

The objective of this work is to compute the scattered field when a plane wave illuminates a  random dielectric medium bounded by a rough surface. From now on, we write the spatial coordinates $\vr = (x,y,z) = (\vx,z)$. The random rough surface is defined by $z=h(\vx)$; the dielectric permittivity has random fluctuations $\varepsilon_1(\vx,z)$ on top of a background value $\epsilon_1$, so that

\begin{equation}
\epsilon(\vx,z) = \epsilon_0\Big\{ 1\,\Theta(z-h(\vx)) + \left[\epsilon_1 + \varepsilon_1(\vx,z) \right]\Theta(h(\vx)-z)\Big\} \equiv \epsilon_0\,\epsilon(\vx,z) \label{ep_r}    
\end{equation}

As we will deal with a configuration that has statistical azimuthal symmetry (that is, correlation functions are invariant for rotation around the $z$-axis-) it is convenient to perform a 2D Fourier transform for any physical quantity $A(\vx,z)$ as follows

\begin{align}
    & A(\vx,z) = \int d^2p\,e^{\imath \vp\cdot\vx}\,a(\vp,z) \label{2D_Fourier}\\ 
    & a(\vp,z) = \int \frac{d^2x}{(2\pi)^2}\,e^{-\imath \vp\cdot\vx}\,A(\vx,z)  \nonumber
\end{align}

where $\vp = (p_x,p_y)$ is the horizontal component of the mode with vertical component $P$. We place $(2\pi)^{-2}$ on the forward transform, and use this convention throughout. A product of two fields transforms as a convolution with no extra factor,

\begin{equation}
 \left(AB\right)(\vp,z)=\int d^2q\;a(\vq,z)\,b(\vp-\vq,z)\,;
 \label{convolution_rule}
\end{equation}

Then, a plane wave transforms as $e^{\imath\vk\cdot\vx}\rightarrow\delta(\vp-\vk)$, and the spectral density of a stationary zero-mean process, defined under the same conventions, appears without additional factors,

\begin{align}
 \left\langle a(\vp,z)\,a^*(\vp',z')\right\rangle &=W_a(\vp,z,z')\,\delta(\vp-\vp'), \nonumber \\
  W_a & =\int\frac{d^2u}{(2\pi)^2}e^{-\imath\vp\cdot\vec{u}}\left\langle A(\vec{u}+\vx,z)A^*(\vx,z')\right\rangle\,.
 \label{spectral_convention}
\end{align}

This is the convention in which the roughness spectrum of Section~\ref{sec:first_order_sources} is written, and the one used in the remote-sensing literature we compare against \cite{tsang2004scattering}. Note that $\delta(\vec 0)=\int d^2x/(2\pi)^2$ is the illuminated area divided by $(2\pi)^2$.

For the electric field, we assume that $\vE(\vr,t) = e^{\imath \omega t}\,\vE(\vx,z)$. $\mu\epsilon_0\omega^2= k^2_i$ is the wavenumber in vacuum. We write the vertical component of the wave number vector as $P^2_n = \epsilon_nk^2_i - p^2$ where $\epsilon_> = 1$ when $z > h(\vx)$ and $\epsilon_< =\epsilon_1$ when $z<h(\vx)$. The vertical component is then a piece-wise function defined by the rough surface: $P^2_n = P^2_0\,\Theta(z-h(\vx)) + P^2_1\,\Theta(h(\vx)-z)$.

As the rough surface does not present any free density charge or current, the electric and magnetic fields satisfy the Helmholtz equation, with nontrivial matching conditions at the random interface $z = h(\vx)$. Our strategy is to map this problem into an equivalent one where the interface is flat but there are dielectric fluctuations both below and above the surface.

We shall now introduce a new coordinate $\zeta(z,\vx)$ such that the interface becomes the $\zeta=0$ plane, and $\zeta>0$ implies $z>h(\vx)$. Therefore, we define 

\begin{align}
     &z = \zeta + h(\vx)\,f(\z) \equiv \z + \gamma(\vx,\z) \label{z_to_Z}  \\
     &f(0)=1\,,\qquad f(\z)\rightarrow0 \ \text{ when }\ \z\rightarrow\pm\infty\,.\label{gauge_conditions}
\end{align}

\begin{figure}[ht]
    \centering
    \includegraphics[width=\linewidth]{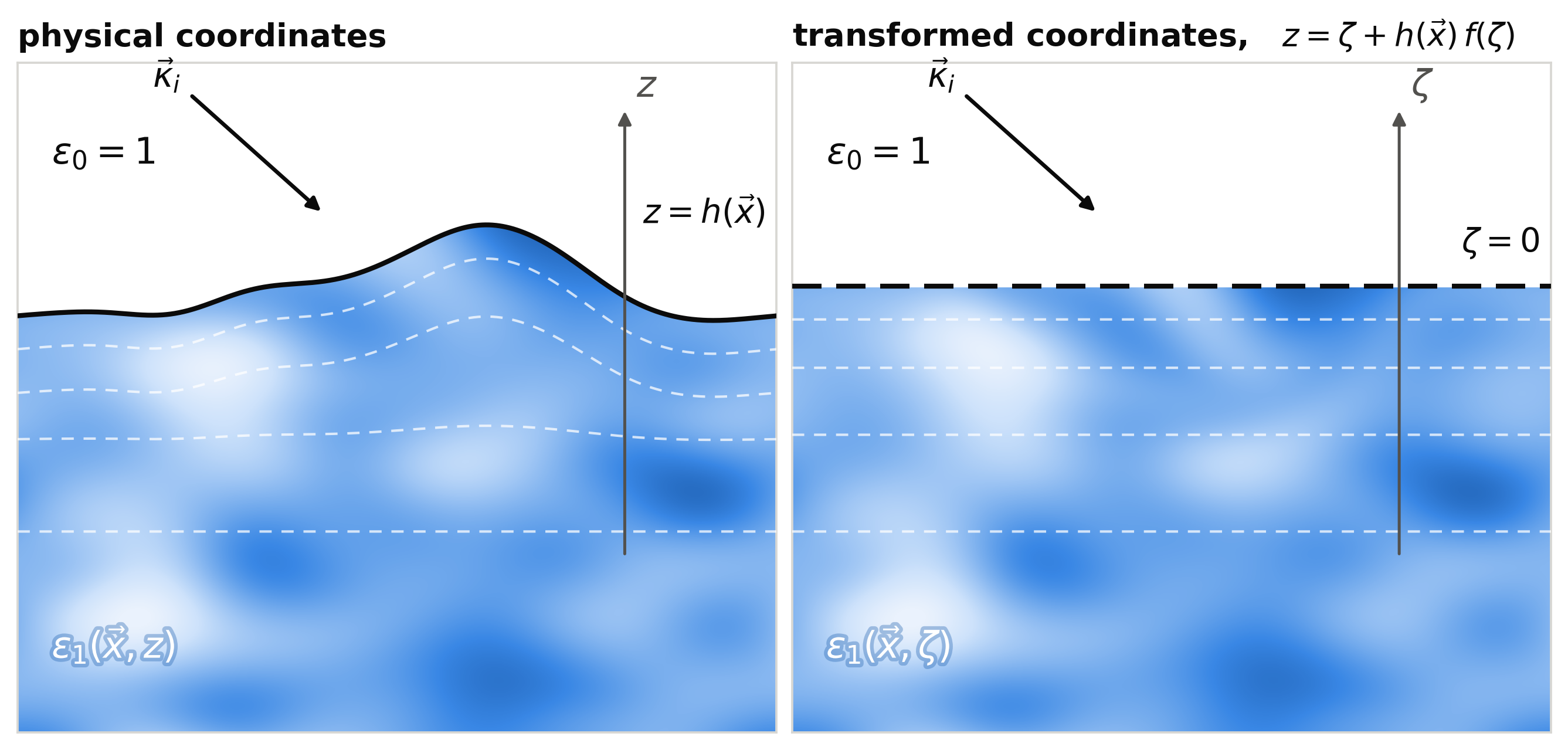}
    \caption{The random rough surface is mapped onto a flat interface by the change of variables (\ref{z_to_Z}). Left: physical coordinates, with the interface $z=h(\vx)$ and, below it, one realization of the inhomogeneous medium $\varepsilon_1(\vx,z)$. Right: the same realization in the transformed coordinates, where the interface is the plane $\z=0$. Dashed curves are surfaces of constant $\z$: they follow the profile near the boundary and flatten with depth, since $f\rightarrow0$.}
    \label{fig:z_to_Z}
\end{figure}

With the new coordinates, the dielectric permittivity results,

\begin{equation}
 \epsilon(\vx,\z) = \epsilon_0\Big\{ 1\,\Theta(\z) + \left[\epsilon_1 + \varepsilon_1(\vx,z) \right]\Theta(-\z)\Big\} \equiv \epsilon_0\,\epsilon_r(\vx,\z)    \label{ep_r_curv}
\end{equation}

\subsection{Change of coordinates, changes in the metric}

The coordinates $(\vx,\zeta)$ define a curvilinear system with metric

\begin{align}
     ds^2 = (dx)^2 + (dy)^2 + \left[\gx\,dx + \gy\,dy + \left(1+\gz\right)\,d\z \right]^2
\end{align}

We see that the metric has components of the first and second order in the derivatives of the function $\gamma(\vx,\z)$. Formally, we collect these terms according to the number of derivatives

\begin{align}
    & \bar{\bar{g}} = \bar{\bar{I}} + \bar{\bar{g}}^{(1)} + \bar{\bar{g}}^{(2)} \label{geometry}
\end{align}

where $\bar{\bar{I}}$ is the identity matrix.  The Jacobian for the transformation defined by (\ref{z_to_Z}) is

\begin{align}
    J = \sqrt{g} = 1 + \gz \label{jacobiano}
\end{align}

We observe that, in curvilinear coordinates, the antisymmetric symbol acquires a factor $1/\sqrt{g}$ \cite{landau2013classical}. Hence, the Helmholtz equation now is,

\begin{equation}
-\epsilon_r(\vx,\z)\epsilon_0\mu\omega^2\,E_a + g_{ab}\frac{1}{\sqrt{g}}e^{bcd}\partial_c\,g_{de}\frac{1}{\sqrt{g}}e^{efg}\partial_fE_g = 0 \label{Helm_curv}    
\end{equation}

So far, we have proceeded in a non-explicit manner: we just applied a general change of coordinates on the Helmholtz equation. But, as we established in (\ref{geometry}), the metric is the identity plus corrections. Then, if we develop it, we can recognize the standard differential operator $k^2_i + \nabla^2$ on the left hand side (LHS) of (\ref{Helm_curv}). The remaining terms may be shifted to the right hand side of the equation, and treated as sources.  As we show below, the sources of the equation depend on the same unknown fields that appear on the LHS. A common approach to solving this kind of equations is to develop a perturbative scheme in which the sources for the field of order $n$ are given by the fields of order $\leq n-1$.

\subsection{Sources from dielectric and geometric fluctuations}

With the change of coordinates used in (\ref{z_to_Z}) we have shown that the metric has terms of zeroth, first, and second order in $\gamma$. The factor $1/\sqrt{g}$ can also be expanded as a leading order in $\gamma$ plus corrections (proportional to the powers in $\gz$). Explicitly, we write

\begin{equation}
\frac{1}{\sqrt{g}}g_{ab} = \delta_{ab} + G_{ab} \label{Gab_def}    
\end{equation}

and use it in (\ref{Helm_curv}) to obtain the usual Helmholtz differential operator on the LHS:

\begin{align}
    &-k^2_i\epsilon_r(\vx,\z)\,E_a + \delta_{ab}e^{bcd}\partial_c\delta_{de}e^{efg}\partial_fE_g = -M_a(\vx,\z),\quad a = \{x,y,\z\}\label{Helm_sources}\\
    & \nn
    & M_a(\vx,\z) = \delta_{ab}e^{bcd}\partial_c G_{de}e^{efg}\partial_fE_g + G_{ab}e^{bcd}\partial_c\left(\delta_{de}+G_{de}\right)e^{efg}\partial_fE_g \label{geom_sources}
\end{align}

Hence, in (\ref{Helm_sources}) we have the Helmholtz operator acting on $E_a$, and on the right hand side (RHS) are the sources for this component of the electric field, which come from the change of coordinates defined in (\ref{geom_sources}).

As we are looking for solutions with a specific direction of propagation, it is more convenient to apply the 2D Fourier transform defined in (\ref{2D_Fourier}), and work in momentum space. Here, it is appropriate to remark that the dielectric permittivity is a piece-wise function that has random fluctuations in the $\z<0$ region, which was defined in (\ref{ep_r_curv}). Its Fourier transform reads

\begin{equation}
    \epsilon_r(\vp,\z) = \epsilon_n(\z)\,\delta(\vq) + \varepsilon_1(\vq,\z)\,\Theta(-\z)
\end{equation}
    
Also, in the curvilinear coordinates, the vertical component of the wavenumber is just $P^2_n = P^2_0\,\Theta(\z)+P^2_1\,\Theta(-\z)$. 

To introduce the TE-TM mode decomposition, we write the electric field as \cite{calzetta2024coherent},

\begin{equation}
 \vE(\vx,\z) = \int d^2p\; e^{\imath \vp \cdot\vx}\left[\vap_{\vp}(\z)\,\hat{\z}\times\hat{p} + \psi_{\vp}(\z)\,\hat{p} + E_{\z\,\vp}(\z)\,\hat{\z} \right] \label{Fourier_2D_E}   
\end{equation}

The TE mode is directly related to $\vap$, while the TM mode is related to $\psi$ and $E_\z$. The scalar fields $\vap$ and $\psi$ are related to the 
components of the electric field by means of the following,

\begin{align}
    &\vap_{\vp}(\z) = \frac{1}{p}\left[p_x\,E_{y\,\vp}(\z)-p_y\,E_{x\,\vp}(\z)\right] & E_{x\,\vp}(\z) &= \frac{1}{p}\left[p_x\,\psi_{\vp}(\z)-p_y\,\vap_{\vp}(\z)\right] \label{vap_Ex} \\
    &\psi_{\vp}(\z) = \frac{1}{p}\left[p_x\,E_{x\,\vp}(\z)+p_y\,E_{y\,\vp}(\z)\right] & E_{y\,\vp}(\z) &= \frac{1}{p}\left[p_x\,\vap_{\vp}(\z)+p_y\,\psi_{\vp}(\z)\right] \label{psi_Ey}
\end{align}

Next, the equations for the $\{x,y,\z\}$ components in the Fourier space, obtained through (\ref{Helm_curv}), are converted into equations for the $\vap, \psi$ and $E_\z$. We recall that, in 2D Fourier space, $\frac{\partial}{\partial x} \rightarrow \imath p_x$, and the same holds for $y$. The goal is to decouple each field in order to give a closed equation for one of them. Before moving on to writing the equations, it is worth noting that, in transforming (\ref{Helm_curv}) into 2D Fourier space, we must work with three types of terms,

\begin{align}
    & \partial_a\,E_b(\vx,\z) \rightarrow \int d^2p\; e^{\imath \vx\cdot\vp}\imath\,p_a\,E_b(\vp,\z) \\
    & \partial_\z\,E_b(\vx,\z) \rightarrow \int d^2p\; e^{\imath \vx\cdot\vp}\,E'_b(\vp,\z) \\
    & k^2_i\,\epsilon_r(\vx,\z)\,E_b(\vx,\z) \rightarrow k^2_i\int d^2p\,d^2q\,e^{\imath (\vp+\vq)\cdot\vx}\,\epsilon_r(\vp,\z)\,E_b(\vq,\z)
\end{align}

The strategy is as follows: we first perform the 2D Fourier transform in each curvilinear coordinate of (\ref{Helm_curv}); then, we use (\ref{vap_Ex}) and (\ref{psi_Ey}) to rewrite each equation in terms of $\vap_{\vp}(\z)$ and $\psi_{\vp}(\z)$. It is worth noting that, in the Fourier space, each $E_{a\,\vp}$ equation will have two terms proportional to such a field: one proportional to the constant background permittivity $\epsilon_n$ and an interaction term with the convolution between the field and the dielectric fluctuations below the rough interface.

\subsubsection{Equation for \texorpdfstring{$E_{\z\,\vp}$}{Ez}}

In Fourier space, the equation for $E_{\z\,\vp}$  reads,

\begin{align}
    \epsilon_n k^2_i E_{\z\,\vp}(\z) &+ k^2_i\int d^2q \varepsilon_1(\vp-\vq,\z)\,E_{\z\,\vq}(\z) - p^2E_{\z\,\vp}(\z) \nn 
    &- \frac{\partial}{\partial \z} \left[\imath p_x E_{x\,\vp}(\z) + \imath p_y E_{y\,\vp}(\z)\right] = M_{\z\,\vp}(\z)
\end{align}

Now we use (\ref{psi_Ey}) to eliminate the dependence on $E_x-E_y$. In addition, we use the relation on-shell $\epsilon_n k^2_i = p^2 + P^2_n$ to simplify the two terms proportional to $E_{\z\,\vp}$.  Hence, the vertical component of the electric field obeys an integral equation 

\begin{align}
    &E_{\z\,\vp}(\z) = \imath \frac{p}{ P^2_n}\,\psi'_{\vp}(\z) + \frac{1}{ P^2_n}M_{\z\,\vp}(\z) - \frac{k^2_i}{ P^2_n}\int d^2q\; \varepsilon_1(\vp-\vq,\z)\,E_{\z\,\vq}(\z) \nn
    &E_{\z\,\vp}(\z) = \imath \frac{p}{ P^2_n}\,\psi'_{\vp}(\z) + \frac{1}{ P^2_n}\left[M_{\z\,\vp}(\z) + N_{\z\,\vp}(\z)\right]\label{Ez_eq}
\end{align}

where $M_{\z\,\vp}(\z)$ is read off (\ref{geom_sources}) and the source produced by the dielectric fluctuations is simply

\begin{align}
    N_{\z\,\vp}(\z) = -k^2_i\int d^2q\;\,\varepsilon_1(\vp-\vq,\z)\,\Theta(-\z)\,E_{\z\,\vq}(\z) \label{Nz_q}
\end{align}

In the leading order, we can use $E_{\z\,\vq} = \imath\frac{q}{Q^2_n}\psi'_{\vq}(\z)$ in the integral term of (\ref{Nz_q}).

\subsubsection{Equation for \texorpdfstring{$\vap_{\vp}$}{varphi}}

The equation for $\vap_{\vp}$ is obtained taking the linear combination of $p_y\,E_x(\vp,\z)-p_x\,E_y(\vp,\z)$, which produces

\begin{align}
    &\left[P^2_n + \partial_\z\partial_\z \right] \vap_{\vp}(\z) = M_{\vap\,\vp}(\z) + N_{\vap\,\vp}(\z) \equiv \mathcal{J}_{\vap\,\vp}(\z) \label{vap_eq}
\end{align}

where we defined

\begin{equation}
 M_{\vap\,\vp}(\z) = \frac{1}{p}\left[p_y\,M_{x\,\vp}(\z)-p_x\,M_{y\,\vp}(\z) \right]   
\end{equation}
 
with $M_{x\,\vp}(\z)$ and $M_{y\,\vp}(\z)$ obtained from (\ref{geom_sources}) are the sources introduced by the change of coordinates and

\begin{align}
    N_{\vap\,\vp}(\z) = -\frac{k^2_i}{p}\int d^2q\;\varepsilon_1(\vp-\vq,\z)\,\Theta(-\z)\,\frac{1}{q} \left[\left(\vp\cdot\vq\right)\,\vap_{\vq}(\z) + \hat{\z}\cdot\left(\vp\times\vq \right)\psi_{\vq}(\z) \right] \label{Nvap_p}
\end{align}

are sources induced by fluctuations in dielectric permittivity \cite{calzetta2024coherent}. We note that (\ref{vap_eq}) is also an integral equation due to $N_{\vap}$; therefore, to obtain a closed expression, we must treat it as a perturbative approach: if we seek a solution of order $m$, then the fields appearing in $N_{\vap}$ must be of order $m-1$. The same applies to the integral term in (\ref{Ez_eq}).

\subsubsection{Equation for \texorpdfstring{$\psi_{\vp}$}{psi}}

To get the equation for $\psi_{\vp}$ we must take the combination $p_x\,E_x(\vp,\z)+p_y\,E_y(\vp,\z)$,

\begin{align}
    &\left[\epsilon_n k^2_i + \partial_\z\partial_\z \right]\psi_{\vp}(\z) -\imath p E'_{\z\,\vp}(\z) = M_{\psi\,\vp}(\z) + N_{\psi\,\vp}(\z) \label{psi_1}
\end{align}

In the RHS, we have a similar scheme as in (\ref{vap_eq}); it is straightforward to rewrite it and show that 

\begin{align}
    M_{\psi\,\vp}(\z) & = \frac{1}{p}\left[p_x\,M_{x\,\vp}(\z)+p_y\,M_{y\,\vp}(\z)\right] \nn
    N_{\psi\,\vp}(\z) & = -\frac{k^2_i}{p}\int d^2q\;\varepsilon_1(\vp-\vq,\z)\,\Theta(-\z)\frac{1}{q}\left[\left(\vp\times\vq\right)\cdot\hat{\z}\,\vap_{\vq}(\z) + \left(\vp\cdot\vq\right)\,\psi_{\vq}(\z) \right] \label{Npsi_p}
\end{align}

To get a closed equation for $\psi_{\vp}$ we use the normal derivative of (\ref{Ez_eq}),

\begin{equation}
    E'_{\z\,\vp}(\z) = \imath\frac{p}{P^2_n}\psi''_{\vp}(\z) + \left(\imath\frac{p}{P^2_n}\right)'\psi'_{\vp}(\z) + \frac{d}{d\z}\left(\frac{1}{P^2_n}\left[M_{\z\,\vp}(\z)+N_{\z\,\vp}(\z)\right]\right)
\end{equation}

Therefore, equation (\ref{psi_1}) becomes,

\begin{align}
& \epsilon_n p^2\,\psi_{\vp}(\z) + \left(\epsilon_n\frac{p^2}{P^2_n} \psi'_{\vp}(\z)\right)' \equiv \mathcal{J}_{\psi\,\vp} \label{psi_eq}
\end{align}

where 

\begin{align}
\mathcal{J}_{\psi\,\vp} & = \frac{p^2}{k^2_i}\left\lbrace M_{\psi\,\vp}(\z) + N_{\psi\,\vp}(\z) + \imath\frac{d}{d\z}\left(\frac{p}{P^2_n}\left[M_{\z\,\vp}(\z)+N_{\z\,\vp}(\z)\right]\right) \right\rbrace 
\label{psi_eq_source}
\end{align}

\section{Perturbative expansion} \label{sec:pert_develop}

So far, we have transformed the original problem into integral equations for $E_a$ or, equivalently, for $E_{\z\,\vp}(\z)$ and the scalar fields $\vap_{\vp}(\z)$ and $\psi_{\vp}(\z)$. We now shall present a strategy for the solution of these equations based on perturbation theory.

\subsection{Solution to order \textit{n}} \label{Order_n_sol}

A systematic approach to handle this kind of integral equations is to develop a perturbative scheme \cite{tsang2004scattering,elfouhaily2004critical,ulaby1986microwave}. Any unknown field $\chi_{\vp}$ will be expanded at order $(n)$: $\chi_{\vp}(\z) = \chi^{(0)}_{\vp}(\z) + \chi^{(1)}_{\vp}(\z) + \cdots$, where the zeroth-order term corresponds to the solution of a flat interface separating two homogeneous dielectric media (i.e, proportional to the incident wave and Fresnel coefficients). Then the nth order terms in the sources $\mJ$ are written in terms of $(n-1)$ or lower terms in the expansion of the scalar fields. 

Once the  currents are determined, the solution to any order $(n)$ is written as a superposition of plane waves, where the amplitude of each mode is weighted by the sources:

\begin{align}
     &\chi^{(n)}(\vx,\z) = \int d^2p\,\,e^{\imath \vp\cdot\vx}\,F^\Upa_P(\z)\,\chi^{(n)}_{\vp}\label{plane_waves_exp} \\
    &\chi^{(n)}_{\vp} = \int d\z'\,F^\Dwa_P(\z')\,\int \frac{d^2x'}{(2\pi)^2} e^{-\imath \vp\cdot\vx'}\,\mJ^{(n)}_{\chi}(\vx',\z') \equiv \int d\z'\,F^\Dwa_P(\z')\,\mJ^{(n)}_{\chi\,\vp}(\z') \label{amplitude_plane_waves}
\end{align}

The functions $F^\Upa_P(\z)$ and $F^\Dwa_P(\z)$ depend on the incident and scattered mode. For the TE case, they are defined as,

\begin{align}
    & F^\Upa_{\rm{TE}\,P}(\z) = \frac{1}{\sqrt{2P_0}}\,e^{\imath P_0 \z}\,\,\text{, if }\z >0 \nn 
    & F^\Dwa_{\rm{TE}\,P}(\z) = \frac{1}{\sqrt{2P_0}}
    \begin{cases}
     \,\left[R_H\,e^{\imath P_0 \z} + e^{-\imath P_0 \z} \right] & \text{, if } \z > 0 \nn     
     T_H\,e^{-\imath P_1 \z} & \text{, if } \z < 0
    \end{cases} &
\end{align}

being $R_H = \frac{K_0-K_1}{K_0+K_1}$ and $T_H = 1 + R_H = \frac{2K_0}{K_0+K_1}$ the usual Fresnel reflection and transmission coefficients; $K_0 = k_i\sqrt{1-\sin^2\theta_i}$ and $K_1 = k_i\sqrt{\epsilon_1-\sin^2\theta_i}$ are the vertical component of the wavenumber in the upper and lower dielectric medium. 

For the TM case, we have,

\begin{align}
    & F^\Upa_{\rm{TM}\,P}(\z) = -\imath\frac{P_0}{2p^2}T_V\,e^{\imath P_0 \z}\,\,\text{, if }\z >0 \nn 
    & F^\Dwa_{\rm{TM}\,P}(\z) = 
    \begin{cases}
     \frac{1}{T_V}\,\left[-R_V\,e^{\imath P_0 \z} + e^{-\imath P_0 \z} \right] & \text{, if } \z > 0 \nn     
     e^{-\imath P_1 \z} & \text{, if } \z < 0
    \end{cases} &
\end{align}

being $R_V= \frac{\epsilon_1\,K_0-K_1}{\epsilon_1\,K_0+K_1}$ and $T_V=1-R_V= \frac{2\,K_1}{\epsilon_1\,K_0+K_1}$ the reflection and transmission coefficients for the electric field.

In both cases, we can arrange that $F^\Upa_{\rm m\,P}(\z) = \beta(P)e^{\imath P\z}$, this normalization factor, $\beta(P)$, enforces the unit Wronskian condition ($|\mathcal{W}_m| = 1$) for $m = \{\rm{TE},\rm{TM}\}$,

\begin{align}
    \mathcal{W}_{\rm m} = \left(F^\Upa_{\rm m}\,F^{\Dwa'}_{\rm m}-F^{\Upa'}_{\rm m}F^\Dwa_{\rm m}\right) \Big\vert_{\z=0},\qquad\mathcal{W}_{\rm TE} = \imath,\quad\mathcal{W}_{\rm TM} = 1.
\end{align}

We have used the fact that in the relevant case the observation point $\z$ is always above the source point $\z'$. For a high enough observation point, $\z \approx z$ and $(\vx,z)\cdot(\vp,P)\gg1$, the phase in (\ref{plane_waves_exp}) can be approximated using the stationary phase method, producing \cite{felsen1994radiation,born2013principles},

\begin{equation}
 \chi^{(n)}(\vx,z) \approx - 2\pi\imath \,P_0\,\beta(P)\,\frac{e^{\imath k_i r}}{r}\,\chi^{(n)}_{\vp} \label{far_field_approx}
\end{equation}

where $\theta_s$ is the direction of the observation point and $P_0 = k_i\,\cos(\theta_s)$ is evaluated at the stationary-phase point $\vp = k_i\sin(\theta_s)\,\hat{p}$. The pre-factor follows from the Weyl identity $e^{\imath k_ir}/r = (\imath/2\pi)\int d^2p\,P^{-1}_0e^{\imath(\vp\cdot\vx + P_0\vert z\vert)}$, which is written in the same convention as (\ref{plane_waves_exp}).

As we can see from (\ref{far_field_approx}), the energy flux in the direction $\vp$ depends on the amplitude of $\chi^{(n)}_{\vp}$. This amplitude is a function of the source $\mJ_{\vp}$, which in turns depends on the stochastic variables $h(\vx),\,\varepsilon_1(\vx,\z)$. We are assuming that both stochastic processes are Gaussian, have zero mean and are stationary, therefore, the mean field of all odd-order terms will vanish, while the even-terms will be non-zero only in the specular direction \cite{tsang2004scattering}. This is known as the coherent term. As we are interested in the backscattering case, we must compute the incoherent component of the scattering amplitude, which is defined as \cite{tsang2004scattering,johnson1999third}:

\begin{align}
    \langle |\chi^{(n)}_{\vp}|^2 \rangle = &\int \frac{d^2x_1}{(2\pi)^2}\int \frac{d^2x_2}{(2\pi)^2} e^{-\imath \vp\cdot(\vx_1-\vx_2)} \nn 
    &\int d\z_1\int d\z_2 \,\,F^\Dwa_P(\z_1)\,F^{\Dwa*}_P(\z_2)\,\langle \mJ^{(n)}_{\chi}(\vx_1,\z_1)\,\mJ^{(n)*}_{\chi}(\vx_2,\z_2) \rangle \label{incoh_field}
\end{align}

The mean value of the product between the two sources will be based on defining the kind of autocorrelation of each random variable and, if there is one, the correlation between the roughness and the fluctuations of the dielectric medium.

\begin{align}
    \langle \mJ^{(n)}_{\chi}(\vx_1,\z_1)\,\mJ^{(n)*}_{\chi}(\vx_2,\z_2) \rangle =& \langle M_{\chi}(\vx_1,\z_1)\,M^*_{\chi}(\vx_2,\z_2)\rangle + 
    \langle N_{\chi}(\vx_1,\z_1)\,N^*_{\chi}(\vx_2,\z_2)\rangle \nn
    & \langle M_{\chi}(\vx_1,\z_1)\,N^*_{\chi}(\vx_2,\z_2)\rangle +
    \langle N_{\chi}(\vx_1,\z_1)\,M^*_{\chi}(\vx_2,\z_2)\rangle
\end{align}

As any perturbative scheme, we need to give the zeroth-order solution. In this case, as we mentioned above, the plane wave reflected and transmitted is characterized by a flat interface separating two homogeneous dielectric mediums. 

This solution depends on the polarization of the incident electric field. We assume an incident plane wave that, in Cartesian coordinates, is written as
$\vec{E}(\vx,z) = \hat{p}\,\exp\{\imath (\vk\cdot\vx - K_0 z)\}$, where $\hat{p}$ is the polarization vector in TE or TM mode. Besides the polarization, we can always choose that the incident wavenumber vector $\vka = \vk - K_0 \hat{z}$ has a horizontal component $\vk = k\,\hat{x} = k_i\,\sin(\theta_i)\,\hat{x}$. This choice is justified by the fact that both the dielectric fluctuations and the rough surface present azimuthal symmetry: the result is invariant under a rotation around the vertical axis. Therefore, if the incident wave is in TE mode $\hat{p} = \hat{y}$ and, if it is TM mode, $\hat{p} = \hat{x}+(k/K_0)\,\hat{\z}$.

\subsection{Zeroth-order solution - incident field}

Here we write the lower level fields that will later be used as seeds to obtain the higher contributions in each solution.

For the TE case, we have

\begin{align}
   & \vE^{(0)}(\vx,\z) = \hat{y}\,e^{\imath \vk\cdot\vx}
  \begin{cases} 
   R_H\,e^{\imath K_0\z} + e^{-\imath K_0\z} & \text{if } \z > 0 \\
   T_H\,e^{-\imath K_1\z}      & \text{if } \z < 0
  \end{cases} \label{vap0_TE}
\end{align}

From here, it is immediate to see that $E^{(0)}_{\vp} \sim \delta(\vp-\vk)$. As $k_y = 0$ and then $k_x = k$, following the definitions for the scalar fields, we found that $\vap^{(0)}_{\vp} = E^{(0)}_{y\,\vp}$, meanwhile $\psi^{(0)}_{\vp} = E^{(0)}_{\z\,\vp} = 0$. In this case, the polarization vector remains invariant, $\hat{p} = \hat{\tilde{p}} = \hat{y}$.

With the TM case $E^{(0)}_y = 0$ and as $k_y = 0$, $\vap^{(0)}_{\vp}= 0$. As in this case $\hat{p} = \hat{x}\times\hat{\kappa}$ then we associate $E^{(0)}_x = \psi^{(0)}$, 

\begin{align}
 \psi^{(0)}(\vx,\z) = e^{\imath \vk\cdot\vx}
 \begin{cases}
     -R_V\,e^{\imath K_0 \z} + e^{-\imath K_0\z} &\text{if } \z > 0 \\
     T_V \,e^{-\imath K_1 \z} & \text{if } \z < 0
 \end{cases}   \label{psi0_TM}
\end{align}

and obtain the $E^{(0)}_{\z}$ amplitude using the condition $\vka\cdot\vE = 0$ for both the incident and reflected/transmitted fields. So, for both we have

\begin{align}
   & E^{(0)}_{\z}(\vx,\z) = e^{\imath \vk\cdot\vx}
  \begin{cases} 
   \frac{k}{K_0}\,\left[R_V\,e^{\imath K_0\z} + e^{-\imath K_0\z}\right] & \text{if } \z > 0 \\
   \frac{k}{K_1}\,T_V\,e^{-\imath K_1\z}      & \text{if } \z < 0
  \end{cases} \label{Ez0_TM}
\end{align}

It is straightforward to verify that $\psi^{(0)}$ is continuous at $\z= 0$ but $E^{(0)}_\z$ has a jump due to the continuity of $D^{(0)}_\z$.

Using these fields as primordial sources, we can compute $\mJ^{(n)}$ for $n \geq 1$. In the following Section we list the sources at first order, both for terms 
arising from the metric and those due to fluctuations in the dielectric medium. In both cases, we analyze the TE and TM case.

\section{First order sources} \label{sec:first_order_sources}

As we have discussed above, we need to compute the sources at first order. The fields are kept at zeroth order and the perturbation comes from the curved metric and from the dielectric fluctuations, so that we deal with terms of the form $\mathcal{G}^{(1)}\,\chi^{(0)}$, where $\mathcal{G}^{(1)}$ denotes any contribution of first order in the metric. At first order in metric perturbations,

\begin{align}
    M^{(1)}_a(\vx,\z) &= \left(\delta_{ab}\,G^{(1)}_{de}+G^{(1)}_{ab}\,\delta_{de}\right)\,e^{bcd}e^{efg}\left(\partial_c \partial_f E^{(0)}_g\right) \nn &+\delta_{ab}e^{bcd}e^{efg}\left(\partial_c G^{(1)}_{de}\right)\left(\partial_f E^{(0)}_{g}\right)\label{geom_current_O1}    
\end{align}

To give $G^{(1)}_{ab}$, defined in (\ref{Gab_def}), we must keep in mind two aspects: (i) the metric is written as the sum of the identity and a first order contribution; (ii) the Jacobian of the change of coordinates is $\sqrt{g} = 1 + \gz$ and therefore $1/\sqrt{g}$ has contributions to all orders. Then, it is necessary to develop it to first order in $\gz$. Following this, we compute each of them for the TE or TM incident case. As we can see from (\ref{incoh_field}), the sources appear in momentum space. Then it is convenient to perform the 2D Fourier transform of the rough surface profile, 

\begin{equation}
    H(\vp) = \int \frac{d^2x}{(2\pi)^2}\,e^{-\imath \vp\cdot\vx}\,h(\vx).
\end{equation}

\subsection{TE incidence}

In this case, we use $E^{(0)}_y(\vx,\z)=\vap^{(0)}(\vx,\z)$, defined in (\ref{vap0_TE}). Also,$\psi^{(0)} = E^{(0)}_\z = 0$. In momentum space, the incident field $\vap^{(0)} (\vx,\z)$ results,

\begin{align}
    & \vap^{(0)}_{\vp}(\z) = \delta(\vp-\vk)\,\Phi^{(0)}_{\vk}(\z)\nn 
    & \Phi^{(0)}_{\vk}(\z) = 
     \begin{cases}
         R_H\,e^{\imath K_0 \z} + e^{-\imath K_0 \z} &\text{if } \z>0 \nn 
         T_H\,e^{-\imath K_1 \z} &\text{if } \z <0 \nonumber
     \end{cases} \nonumber
\end{align}

\subsubsection*{TE geometric sources}

The equations in momentum space are,

\begin{align}
    M^{(1)}_{TE\,x\,\vp}(\z) & = p_x\,p_y\,\Phi^{(0)}_{\vk,\z}(\z)\,f(\z)\,H(\vp-\vk) \label{M1x_TE_p} \\
    M^{(1)}_{TE\,y\,\vp}(\z) & = \left\lbrace \left[(k^2-p^2_x)f(\z) -f''(\z) \right]\Phi^{(0)}_{\vk,\z}(\z)
    +2f'(\z)\Phi^{(0)}_{\vk,\z\z}(\z)\right\rbrace\,H(\vp-\vk) \label{M1y_TE_p} \\
    M^{(1)}_{TE\,z\,\vp}(\z) &= -\imath p_y\left[\Phi^{(0)}_{\vk,\z\z}(\z)f(\z)+\Phi^{(0)}_{\vk,\z}(\z)f'(\z)\right]\,H(\vp-\vk)\label{M1z_TE_p}
\end{align}

We rewrite each of them as $M^{(1)}_{TE\,a\,\vp}(\z) = \tilde{m}^{(1)}_{TE\,a}(\vp,\z)\,H(\vp-\vk)$. The sources are bilinear in $h$ and in the zeroth-order field; by the convolution rule (\ref{convolution_rule}) and the transform of a plane wave, the two combine into the single factor $H(\vp-\vk)$ with no factor of $2\pi$ left over.

\subsubsection*{TE dielectric sources}

In this case, the sources that involve dielectric fluctuations are given just by (\ref{Nvap_p}); there we must use $\psi^{(0)}_{\vp} = 0$ and $\vap^{(0)}_{\vp}$ just as before.

\subsection{TM incidence}

Here the pair $(E^{(0)}_\z, \psi^{(0)})$ defined in (\ref{Ez0_TM})-(\ref{psi0_TM}) act as sources and $E^{(0)}_y = 0$. For later convenience, we define an auxiliary field 

\begin{equation}
\mathcal{H}^{(0)}(\vx,\z) =   \psi^{(0)}_{,\z}(\vx,\z) - E^{(0)}_{\z,x}(\vx,\z) \label{aux0_TM}    
\end{equation}

which is proportional to the incident magnetic field $H^{(0)}_y$. This field and its Fourier transform result in

\begin{align}
    &\mathcal{H}^{(0)}(\vx,\z) = e^{\imath \vk\cdot\vx}\,\Psi^{(0)}_{\vk}(\z) \nn
    &\mathcal{H}^{(0)}_{\vp}(\z) = \delta(\vp-\vk)\,\Psi^{(0)}_{\vk}(\z) \nn 
    &\Psi^{(0)}_{\vk}(\z) = -\imath\,k^2_i
     \begin{cases}
         \frac{1}{K_0}\left[R_V\,e^{\imath K_0 \z} + e^{-\imath K_0 \z} \right]
         &\text{if } \z>0 \nn 
         \frac{\epsilon_1}{K_1}T_V\,e^{-\imath K_1 \z} &\text{if } \z <0 
     \end{cases} 
\end{align}

This field is a piece-wise function (as $\vap^{(0)}$ for the TE case), it is continuous at the interface $\z = 0$, but the normal derivative shows a jump. 

\subsubsection*{TM geometric sources}

The corresponding expansion for each source result

\begin{align}
    m^{(1)}_{TM\,x\,\vp}(\z) &= \left\lbrace \left[\left(k(k-p_x)-p_y^2\right)f(\z)+f''(\z)\right]\,\Psi^{(0)}_{\vk}(\z)+2f'(\z)\Psi^{(0)}_{\vk,\z}(\z)\right\rbrace\,H(\vp-\vk) \label{M1x_TM_p}\\
    m^{(1)}_{TM\,y\,\vp}(\z) &= p_y(p_x-k)\,\Psi^{(0)}_{\vk}(\z)\,f(\z)\,H(\vp-\vk) \label{M1y_TM_p}\\
    m^{(1)}_{TM\,\z}(\vp,\z) &= -\imath\,(p_x-k)\left[\Psi^{(0)}_{\vk,\z}(\z)f(\z)+\Psi^{(0)}_{\vk}(\z)f'(\z)\right]\,H(\vp-\vk)\label{M1z_TM_p}
\end{align}

and again, we write them as  $m^{(1)}_{TM\,a}(\vp,\z) = \tilde{m}^{(1)}_{TM\,a\,\vp}(\z)\,H(\vp-\vk)$.

\subsubsection*{TM dielectric sources} 

In this case, the sources that involve dielectric fluctuations are given just by (\ref{Npsi_p}); there we must use $\vap^{(0)}_{\vp} = 0$.

\subsection{Roughness spectrum} \label{subsec:rough_spec}

Now we go back to compute explicit expressions for the mean scattered fields. In order to do so, we first characterize the random rough surface and the fluctuations in the dielectric through their autocorrelation functions and the possible correlation between the two kind of random variables.

We consider the rough surface as a gaussian isotropic process with zero mean and autocorrelation function $C_h$ defined by

\begin{equation}
    s^2\,C_h(\vx-\vx') \equiv \langle h(\vx)\,h(\vx') \rangle     
\end{equation}

where $l$ it is the correlation length of the surface and $s$ its RMS height. As we are interested in observing the amplitude of the scattered field in a certain direction $\vp$ it is convenient to compute the 2D Fourier transform of the autocorrelation function. This is known as the roughness spectrum of the rough surface and reads,

\begin{align}
    s^2\,C_h(\vx-\vx') & = \int d^2p\;e^{\imath \vp\cdot(\vx-\vx')}\,W_h(\vp)\nn
    \left\langle H(\vp)\,H^*(\vp')\right\rangle & = W_h(\vp)\,\delta(\vp-\vp')
\end{align}

in the convention (\ref{2D_Fourier})--(\ref{spectral_convention}).

For a gaussian autocorrelation function, the roughness spectrum is also gaussian \cite{tsang2004scattering}. In the following we assume the above and use,

\begin{equation}
 W_h(\vp) = \frac{s^2\,l^2}{4\pi}\,e^{-l^2\vp^2/4} \label{rough_corr}    
\end{equation}

\subsection{Dielectric fluctuations} \label{subsec:diel_spec}

As explained in the Introduction, we assume that in the curvilinear coordinate system the medium is statistically homogeneous. Moreover, following \cite{zuniga1979active}, we propose that the dielectric fluctuations have a two-fold dependency in the correlation function: gaussian laterally and exponential vertically. Therefore,

\begin{align}
 & C_\vep(\vx-\vx',\z-\z') \equiv \langle \varepsilon_1(\vx,\z)\, \varepsilon^*_1(\vx',\z' )\rangle = s^2_\vep\, e^{-(\vx-\vx')^2/l^2_r}\,e^{-|\z-\z'|/l_v} \nn
 & W_\vep(\vp,\z-\z') = \frac{s^2_\vep\,l^2_r}{4\pi} e^{-l^2_r\vp^2/4}\,e^{-|\z-\z'|/l_v} \label{diel_corr}
\end{align}

where $s_\vep$ indicates the intensity of the fluctuations and $l_r$ ($l_v$) is the lateral (vertical) correlation length. As in (\ref{spectral_convention}), $\left\langle \varepsilon_1(\vp,\z_1)\,\varepsilon^*_1(\vp',\z_2)\right\rangle = W_\vep(\vp,\z_1-\z_2)\,\delta(\vp-\vp')$.

\subsection{Roughness-dielectric correlation} \label{subsec:interc_spec}

A correlation between the dielectric constant fluctuations (volume scattering) and the surface roughness (surface scattering) should be expected when the fluctuations in the material properties are physically coupled to the changes in the surface elevation. This coupling is particularly common in natural media like soil, snow, or ice, where the surface geometry and the subsurface material properties are often driven by the same physical processes.

If both the roughness and dielectric variations arise from a common process (e.g., moisture variation or textural layering), then the statistical dependence between $h(\vx)$ and $\varepsilon_1(\vx,\z)$ decays with depth and lateral separation. We therefore model this correlation as a separable decaying function in both lateral and vertical directions.

A convenient and widely used form \cite{sarabandi2002electromagnetic,kuo2007},\cite{elson1984theory} is

\begin{align}
     \equiv C_{h\,\vep}(\vx-\vx',\z) &\equiv \langle h(\vx)\,\varepsilon^*_1(\vx',\z')\rangle = \rho_0\,s\,s_\vep\,e^{-(\vx-\vx')^2/L^2_r}\,e^{-|\z'|/L_v} \nn 
      W_{h\vep}(\vp,\z') &= \rho_0\,\frac{s_\vep\,s\,L^2_r}{4\pi}e^{-L^2_r p^2/4}\,e^{-|\z'|/L_v} \label{h_vep_corr}
\end{align}

where $\rho_0 \in [-1,1]$ is the surface-dielectric correlation coefficient, measuring the degree of coupling between the two stochastic variables $H(\vp)$ and $\varepsilon_1(\vp',\z')$. 

\section{Mean values}\label{sec:mean_values}

In this Section, we compute the mean scattered field that propagates in a general direction $\vp$. later we will specify $\vp=-\vk$, that is, the backscattering condition, where $p_x=-k_x$, $p_y = 0$, and then $p = k = |k_x|$. As we assume both the rough surface and the dielectric fluctuations are Gaussian processes with zero mean, the mean value of the scattered field will be non-zero only in the specular direction. This is known as the coherent component \cite{tsang2004scattering}. As  we are interested in the backscattering condition, we need to compute the incoherent component of the scattered field, which was formally defined in (\ref{incoh_field}).

\subsection{TE-TE mode}

In what follows, we develop this for the case where both incident and scattered fields are in TE mode (HH-channel). Therefore, we will use the sources defined in (\ref{Nvap_p}).

We must compute the mean square value of

\begin{align}
  \vap^{(1)}_{TE\,\vp} &= \int d\z'\,F^\Dwa_{\rm{TE}\,P}(\z')\, \frac{1}{p} \left[p_x\,\tilde{m}^{(1)}_{TE\,y}(\vp,\z')-p_y\,\tilde{m}^{(1)}_{TE\,x}(\vp,\z')\right]\,H(\vp-\vk)  \nn
  &-\frac{k^2_i}{p}\int d\z'\,F^\Dwa_{\rm{TE}\,P}(\z')\varepsilon_1(\vp-\vk,\z')\frac{1}{k}(\vp\cdot\vk)\,\Phi^{(0)}_{\vk}(\z')\Theta(-\z')\label{scatt_field_O1}
\end{align}

At this point, we must distinguish whether the source point $\z'$ is located below or above the plane $\z'=0$. Above there are incident and reflected waves; below, at leading order, there are only transmitted waves. Moreover, there are geometric sources in both regions, but dielectric fluctuations exist only below the interface.

Using the correlation functions introduced in Sections~\ref{subsec:rough_spec}-\ref{subsec:interc_spec} and writing $\mathcal{A} \equiv \delta(\vec{0})=\int d^2x/(2\pi)^2$, the illuminated area divided by $(2\pi)^2$ in the convention of (\ref{2D_Fourier}), and specifying (\ref{scatt_field_O1}) at the backscattered direction (i.e. $p_x = - k$, $p_y = 0$), we get,

\begin{align}
     \frac{1}{\mathcal{A}}\langle |\vap^{(1)}_{TE\,-\vk}|^2 \rangle &= W_h(\vp-\vk)\,\Big\vert\int^\infty_{-\infty}d\z_1\,F^\Dwa_{\rm{TE}\,P}(\z_1)\,
     \tilde{m}^{(1)}_{TE\,y}(-\vk,\z_1)\Big\vert^2\nn
     & + k^4_i\,\frac{(-\vk\cdot\vk)^2}{(p\,k)^2} \,\iint^0_{-\infty} d\z_1\,d\z_2\,W_\vep(-2\vk,\z_1-\z_2)\,F^\Dwa_{\rm{TE}\,P}(\z_1)\,F^{*\Dwa}_{\rm{TE}\,P}(\z_2)\Phi^{(0)}_{\vk}(\z_1)\,\Phi^{*(0)}_{\vk}(\z_2) \nn
     &-2\Re\left\lbrace \frac{k^2_i}{p}\,\int^\infty_{-\infty} d\z_1\,F^\Dwa_{\rm{TE}\,P}(\z_1)\,\left[-\tilde{m}^{(1)}_{TE\,y}(-\vk,\z_1)\right]\right. \nn
     &\left.\hspace{2cm}\frac{(-\vk\cdot\vk)}{k}\,\int^0_{-\infty} d\z_2\,W_{h\vep}(-2\vk,\z_2)\,F^{*\Dwa}_{\rm{TE}\,P}(\z_2)\,\Phi^{*(0)}_{\vk}(\z_2) \right\rbrace \label{HH_meansquare}
\end{align}

Now we need to write the sources using (\ref{M1x_TE_p}), (\ref{M1y_TE_p}), use (\ref{Nvap_p}) and compute the integrals. The first term is the geometric one, and in backscattering condition it gives the scattering amplitude of SPM \cite{tsang2004scattering,johnson1999third}; its evaluation is carried out in full in the accompanying repository \cite{franco2026code} and we quote the result below. The second and third terms reduce to the two $\z$-integrals of Appendix~\ref{appendix_inegrals}, where they are computed for both channels, the second with (\ref{diel_corr}) and the third with (\ref{h_vep_corr}). Hence, in backscattering condition, we find,

\begin{align}
    \frac{1}{\mathcal{A}}\langle |\vap^{(1)}_{TE\,-\vk}|^2 \rangle &= \Big\vert I^h_{HH}(-\vk)\Big\vert^2 \frac{s^2l^2}{4\pi}e^{-(kl)^2} \nn
    & + k^4_i \frac{|T_H|^4}{4 K_0\,K''_1}\frac{s^2_{\vep}\,l^2_r}{4\pi}e^{-(kl_r)^2}\,\frac{\tilde{l}_v}{1+\left(2K'_1\tilde{l}_v\right)^2} \nn
    &-2\Re\left\lbrace 2k^2_iK_0R_H(T^*_H)^2\,\rho_0\frac{s_{\vep}\,s\,L^2_r}{4\pi}e^{-(kL_r)^2}\,\frac{L_v}{1+2L_v(K''_1+\imath K'_1)}\right\rbrace\label{HH_O1}
\end{align}

with $\tilde{l}_v=l_v/(1+2K''_1l_v)$ the effective vertical correlation length of
(\ref{eq:Jeps}), and with the geometric amplitude

\begin{equation}
    I^h_{HH}(-\vk) \;=\; -\frac{4R_HK^2_0}{\sqrt{2K_0}}\,,
    \label{IHH_exact}
\end{equation}

the standard Bragg amplitude, proportional to $R_H$, as required by the first-order SPM kernel $\alpha_{HH}=(\epsilon_1-1)/(K_0+K_1)^2=-R_H/k^2_i$ \cite{tsang2004scattering,johnson1999third,ulaby1986microwave}. The detailed development of the algebra is elementary but long  and is reproduced full symbolically in the accompanying repository \cite{franco2026code}.

Here, it is worth noting that if we expect some contribution from the third term, there must be correlations between the dielectric fluctuations and the geometric effects above the interface $\z = 0$. This is due to the fact that in backscattering, the integration of geometric sources in the region $\z<0$ vanishes. Additionally, we note that these results were found without specifying the function $f(\z)$ that defines the change in coordinates $z\rightarrow \z$. So far, the only requirement on the change of coordinates was that in the new one the translational invariance on the correlation function for the dielectric fluctuations.

The main result of our development is the last line in (\ref{HH_O1}). It shows that if there is a  correlation between the roughness and the dielectric fluctuations, this effect produces a new scattering mechanism even at first order. This means that the two random characteristics of the target are not purely additive.

\subsection{TM-TM mode}

In this case, in the backscattering condition, we must compute the mean square value of

\begin{equation}
  \psi^{(1)}_{TM\,-\vk} = I^h_{VV}(-\vk)\,H(-2\vk) \;+\; I^{\vep}_{VV}(-\vk)
  \label{scatt_field_O1_TM}
\end{equation}

which, as in (\ref{amplitude_plane_waves}), is the amplitude of the mode $-\vk$; the field is recovered as $\psi^{(1)}(\vx,\z)\rightarrow F^\Upa_{\rm{TM}\,K}(\z)\,\psi^{(1)}_{TM\,-\vk}$. The two contributions come from the geometric and dielectric parts of the source (\ref{psi_eq_source}),

\begin{align}
  I^h_{VV}(-\vk) &= \frac{k^2}{k^2_i}\int d\z'\,F^\Dwa_{\rm{TM}\,K}(\z')\,\left[-\tilde{m}^{(1)}_{TM\,x}(-\vk,\z')+\imath\frac{d}{d\z'}\frac{k}{K^2_n}\tilde{m}^{(1)}_{TM\,\z}(-\vk,\z') \right] \label{IVV_def} \\
  I^{\vep}_{VV}(-\vk) &= \frac{k^2}{k^2_i}\int d\z'\,F^\Dwa_{\rm{TM}\,K}(\z')\,\left[N^{(1)}_{\psi\,-\vk}(\z')+\imath\frac{d}{d\z'}\frac{k}{K^2_n}N^{(1)}_{\z\,-\vk}(\z') \right] \label{IVVeps_def}
\end{align}

with $d/d\z'$ understood in the distributional sense, so that the discontinuities of $K^2_n$ and of the sources at $\z'=0$ are automatically included. For TM incidence $\vap^{(0)}_{\vk}=0$, so that (\ref{Nz_q}) and (\ref{Npsi_p}) give, at first order and in backscattering,

\begin{align}
    N^{(1)}_{\psi\,-\vk}(\z) &= +k^2_i\,\varepsilon_1(-2\vk,\z)\,\Theta(-\z)\,\psi^{(0)}_{\vk}(\z) \nn
    N^{(1)}_{\z\,-\vk}(\z) &= -k^2_i\,\varepsilon_1(-2\vk,\z)\,\Theta(-\z)\,E^{(0)}_{\z\,\vk}(\z)
    \label{Neps_backscatter}
\end{align}

Both sources are supported in $\z<0$ and vanish as $\z\rightarrow-\infty$, so a single integration by parts of the second term of (\ref{IVVeps_def}) leaves no surface term --- the jump of $\Theta(-\z)$ at the interface is interior to the integral and is accounted for automatically. Hence,

\begin{align}
    I^{\vep}_{VV}(-\vk) & =\int^0_{-\infty} d\z\;\varepsilon_1(-2\vk,\z)\,\mathcal{N}(\z)\nn
    \mathcal{N}(\z) &= k^2\left[F^\Dwa_{\rm{TM}\,K}(\z)\,\psi^{(0)}_{\vk}(\z)
    +\imath\frac{k}{K^2_1}\,F'^{\Dwa}_{\rm{TM}\,K}(\z)\,E^{(0)}_{\z\,\vk}(\z)\right]
    \label{IVVeps}
\end{align}

Below the interface $F^\Dwa_{\rm{TM}\,K}(\z)=e^{-\imath K_1\z}$, $\psi^{(0)}_{\vk}(\z)=T_Ve^{-\imath K_1\z}$ and $E^{(0)}_{\z\,\vk}(\z)=(k/K_1)T_Ve^{-\imath K_1\z}$, so the two terms of $\mathcal{N}$ are $k^2T_V$ and $k^4T_V/K^2_1$ times $e^{-2\imath K_1\z}$, and their sum collapses by means of $K^2_1+k^2=\epsilon_1k^2_i$:

\begin{equation}
    \mathcal{N}(\z)=\frac{k^2\left(K^2_1+k^2\right)}{K^2_1}\,T_V\,e^{-2\imath K_1\z}
    =\frac{\epsilon_1\,k^2_i\,k^2\,T_V}{K^2_1}\,e^{-2\imath K_1\z}\;\equiv\;\mathcal{N}_0\,e^{-2\imath K_1\z}
    \label{Nkernel}
\end{equation}

a single exponential, exactly as in the HH channel. Note that $\mathcal{N}_0$ is proportional to $\epsilon_1$ and not to $\epsilon_1-1$: unlike roughness, the dielectric fluctuations radiate from the bulk of the lower medium and do not require a contrast across the interface. The mean square of (\ref{scatt_field_O1_TM}) has the same three-term structure as the HH channel,

\begin{align}
    \frac{1}{\mathcal{A}}\left\langle \left\vert \psi^{(1)}_{TM\,-\vk}\right\vert^2\right\rangle
    &= \left\vert I^h_{VV}(-\vk)\right\vert^2\,W_h(-2\vk) \nn
    &+ \left\vert\mathcal{N}_0\right\vert^2\iint^0_{-\infty} d\z\,d\z'\;W_{\vep}(-2\vk,\z-\z')\,
    e^{-2\imath\left(K_1\z-K^*_1\z'\right)} \nn
    &+ 2\Re\left\lbrace I^h_{VV}(-\vk)\,\mathcal{N}^*_0\int^0_{-\infty} d\z\;W_{h\vep}(-2\vk,\z)\,e^{2\imath K^*_1\z}\right\rbrace
    \label{VV_meansquare}
\end{align}

The geometric amplitude of (\ref{IVV_def}) results,

\begin{align}
    I^h_{VV}(-\vk) &= \frac{4R_V}{T_V}\,k^2 \;+\; \frac{2\epsilon_1(\epsilon_1-1)\,k^4}{K^2_1}\,T_V
    \;=\;\frac{2k^2\left[\epsilon^2_1K^2_0-K^2_1+2\epsilon_1(\epsilon_1-1)k^2\right]}
    {K_1\left(\epsilon_1K_0+K_1\right)}\,.
    \label{eq:IVV_exact}
\end{align}

which verifies the SPM amplitude \cite{tsang2004scattering,johnson1999third,ulaby1986microwave}. Similarly for the HH channel, the details of the calculation for $I^h_{VV}$ can be verified in \cite{franco2026code}.

The two $\z$-integrals left in (\ref{VV_meansquare}) are those of the Appendix~\ref{appendix_inegrals}, the same ones already met in the HH channel. Therefore, the incoherent backscattered power in the VV channel results,

\begin{align}
    \frac{1}{\mathcal{A}}\left\langle \left\vert \psi^{(1)}_{TM\,-\vk}\right\vert^2\right\rangle
    &= \left\vert I^h_{VV}(-\vk)\right\vert^2\,\frac{s^2\,l^2}{4\pi}e^{-(kl)^2} \nn
    &+ \frac{\left\vert\epsilon_1\right\vert^2k^4_ik^4\left\vert T_V\right\vert^2}{\left\vert K_1\right\vert^4}\,
    \frac{s^2_{\vep}\,l^2_r}{4\pi}e^{-(kl_r)^2}\,\frac{\tilde{l}_v}{1+\left(2K'_1\tilde{l}_v\right)^2}\,\frac{1}{2K''_1} \nn
    &+ 2\Re\left\lbrace I^h_{VV}(-\vk)\,\frac{\epsilon^*_1k^2_ik^2T^*_V}{\left(K^*_1\right)^2}\,
    \rho_0\frac{s_{\vep}\,s\,L^2_r}{4\pi}e^{-(kL_r)^2}\,\frac{L_v}{1+2L_v\left(K''_1+\imath K'_1\right)}\right\rbrace
    \label{VV_O1}
\end{align}

The corresponding field amplitude follows from $E^{(1)}_{VV}=-(k_i/K_0)\,\psi^{(1)}_{TM\,-\vk}$, so that $\langle\vert E^{(1)}_{VV}\vert^2\rangle=(k^2_i/K^2_0)\langle\vert\psi^{(1)}_{TM\,-\vk}\vert^2\rangle$.

As in the HH channel, the last line of (\ref{VV_O1}) is the genuinely new mechanism: a correlation between roughness and dielectric fluctuations contributes to the backscattered power already at first order, so that the two random characteristics of the target are not additive. There is, however, a difference with (\ref{HH_O1}). There the geometric integral over $\z<0$ vanishes in backscattering, so that the cross term could only survive if the dielectric fluctuations were correlated with geometric effects above the interface. Here $I^h_{VV}$ receives contributions from both regions and the cross term survives without that restriction. The VV channel is therefore the more favorable one to look for this effect.

\section{Results} \label{sec:results}

In order to compare the results of Section~\ref{sec:mean_values}, we define the backscattering radar cross-section as the ratio of the scattered power in the far zone to the incident power density, per unit illuminated area $A$ \cite{tsang2004scattering,ulaby1986microwave},

\begin{equation}
    \sigma^0_{QQ} \;\equiv\; \lim_{r\to\infty}\,
    \frac{4\pi\,r^2}{A}\,
    \frac{\left\langle\left\vert E^{(1)}_{s\,QQ}\right\vert^2\right\rangle}
         {\left\vert \vE^{inc}\right\vert^2}\,.
    \label{cross_section}
\end{equation}

Before giving the expression of $\sigma^0$ for both channels, we must clarify three aspects of our development. The first concern to the far-field limit (\ref{far_field_approx}), which gives $r^2\vert E^{(1)}_s\vert^2 = 4\pi^2P^2_0\vert\beta(P)\vert^2\vert\chi^{(1)}_{\vp}\vert^2$,
with $P_0=K_0$ in backscattering. The second point is the amplitude of the incident wave; whereas in TE $\vert\vE^{inc}\vert=1$, it is not unity in the TM channel: the convention (\ref{psi0_TM})--(\ref{Ez0_TM}) fixes $E^{inc}_x=1$ and hence $\vert\vE^{inc}\vert=k_i/K_0$. In VV the factor $k_i/K_0$ introduced by the projection $E^{(1)}_{VV}=-(k_i/K_0)\,\psi^{(1)}$ cancels against it, so that both channels are described by the same expression once one works with the scattered amplitude per unit incident amplitude,

\begin{equation}
    \widehat{E}^{(1)}_{HH}=\frac{1}{\sqrt{2K_0}}\vap^{(1)}_{TE\,-\vk}\,,
    \qquad
    \widehat{E}^{(1)}_{VV}=\frac{K_0\,T_V}{2k^2}\,\psi^{(1)}_{TM\,-\vk}\,,
    \label{Ehat_def}
\end{equation}

which is the normalization for which the geometric amplitudes (\ref{IHH_exact}) and (\ref{eq:IVV_exact}) reproduce the SPM kernels.

The third aspect is related to $\mathcal{A}$. In convention (\ref{2D_Fourier}), $\mathcal{A}\equiv\delta(\vec{0})=\int d^2x/(2\pi)^2$ is the illuminated area divided by $(2\pi)^2$, and not the area $A$ of (\ref{cross_section}) itself: the mean squares (\ref{HH_O1}) and (\ref{VV_O1}) are normalized to $\mathcal{A}=A/(2\pi)^2$.

Collecting the three issues discussed above, the cross-section takes the same form in both channels,

\begin{align}
    \sigma^0_{QQ} &= 4\pi\,K^2_0\,\left\vert\beta(K)\right\vert^2\,\frac{1}{\mathcal{A}}
    \left\langle\left\vert\chi^{(1)}_{-\vk}\right\vert^2\right\rangle \nn
    \sigma^0_{QQ} &= 4\pi\,K^2_0\,\frac{1}{\mathcal{A}}
    \left\langle\left\vert\widehat{E}^{(1)}_{QQ}\right\vert^2\right\rangle\,.
    \label{sigma0_working}
\end{align}

In the roughness-only limit ($s_\vep\to0$) this reproduces the classical first-order cross-section of the Small Perturbation Method \cite{tsang2004scattering,johnson1999third,ulaby1986microwave},

\begin{equation}
 \sigma^0_{QQ}\Big\vert=16\pi\,k^4_i\cos^4\theta\,
    \left\vert\alpha_{QQ}\right\vert^2 W_h(2k)\,,
    \label{spm_sigma0}
\end{equation}

with $\alpha_{HH}$ and $\alpha_{VV}$ given in (\ref{IHH_exact}) and (\ref{eq:IVV_exact}).

For both channels, the radar cross-section given by (\ref{HH_O1}) and (\ref{VV_O1}), is a sum of three terms, which we denote $I_{\rm rough}$, $I_{\rm diel}$ and $I_{\rm cross}$ (their first, second and third lines, respectively). Each quantity plotted in the following is a ratio of these three within one channel, so the results presented do not depend on the global factor in (\ref{sigma0_working}).

Unless stated otherwise, the results shown below use a single parameter set, with all lengths expressed in free-space wavelengths: $\epsilon_1=15+3\imath$, surface r.m.s.\ height $s=0.015\lambda$ and lateral correlation length $l=0.5\lambda$; dielectric fluctuation amplitude $s_\vep=4$ with $l_r=0.5\lambda$ and $l_v=0.2\lambda$; and, for the cross-correlation, $L_r=0.5\lambda$ and $L_v=0.03\lambda$. The roughness is well inside the SPM validity range, $k_i\,s\simeq0.09<0.3$ \cite{ulaby1986microwave}. The scripts that produce every Figure presented in this Section are available in the accompanying repository \cite{franco2026code}.

\subsection{Relative contribution of each scattering mechanism}

\begin{figure}[htbp]
    \centering
    \includegraphics[width=\textwidth]{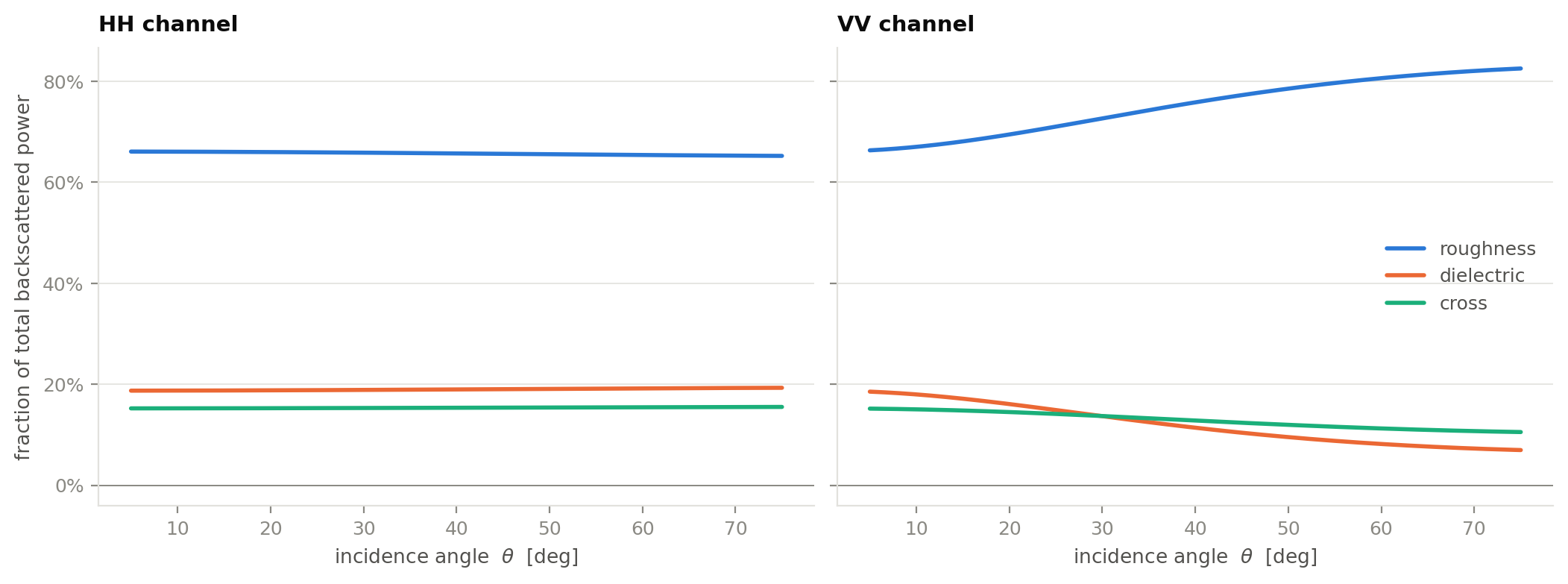}
    \caption{Fraction of the total backscattered power carried by the roughness, dielectric and cross terms, as a function of incidence angle, for $\rho_0=0.5$. In VV the first two have distinct angular signatures and the roughness term dominates over the whole range at these parameters;
    in HH their ratio is fixed by the identity of Fresnel coefficients and the angular sweep is uninformative. The cross term is flat in both channels, and so leaves no shape anomaly that would reveal it.}
    \label{fig:composition}
\end{figure}

In Figure~\ref{fig:composition} we show the fraction of the total backscattered power carried by each of the three terms as a function of the angle of incidence. The two channels present a structural difference. In VV, roughness and volume scattering have clearly distinct angular signatures: while the
roughness fraction rises when the angle of incidence increases, the dielectric one falls by a half for grazing incidence. On the other hand, channel HH presents a quite different behavior. The two mechanisms maintain an essentially fixed ratio over the whole range for the angle of incidence. This is due to the relationship between the reflection and transmission coefficients of the TE mode: $K^4_0\left\vert R_H\right\vert^2 = k^4_i\left\vert\epsilon_1-1\right\vert^2\,\left\vert T_H\right\vert^4/16$. Using this in the first two lines of (\ref{HH_O1}), we get

\begin{equation}
    \frac{I_{\rm rough}}{I_{\rm diel}}\bigg\vert_{HH}
    = 2\left\vert\epsilon_1-1\right\vert^2K''_1\;\frac{W_h}{\Omega_\vep}\,,
    \label{HH_ratio}
\end{equation}

and, when the two lateral correlation lengths coincide, $l=l_r$, the Gaussian factors of $W_h$ and $\Omega_\vep$ cancel out, and the whole angular dependence depends just on the slow variation of $K_1$. Therefore, with the condition $l=l_r$, an angular sweep does not provide any information about the partition between surface and volume scattering for channel HH.

In addition, we observe in Figure \ref{fig:composition} that the cross term is roughly flat in both channels, contributing as much as $15\%$ of the power through the range for the angle of incidence. Hence, with the parameters used, the coupling is not easy to detect: it is just a rescaling factor rather than a structural shape in $\sigma^0(\theta)$ curves.

\subsection{Departure from additivity} \label{sec:balance}

The contribution of the interference between the roughness and dielectric fluctuations can be measured through the error that is produced by ignoring such an interaction. In other words, we measure the failure to assume additivity for the scattering mechanism by

\begin{equation}
    \eta \;\equiv\; \frac{I_{\rm cross}}{I_{\rm rough}+I_{\rm diel}}\,,
    \label{eta_def}
\end{equation}

\begin{figure}[htbp]
    \centering
    \includegraphics[width=\textwidth]{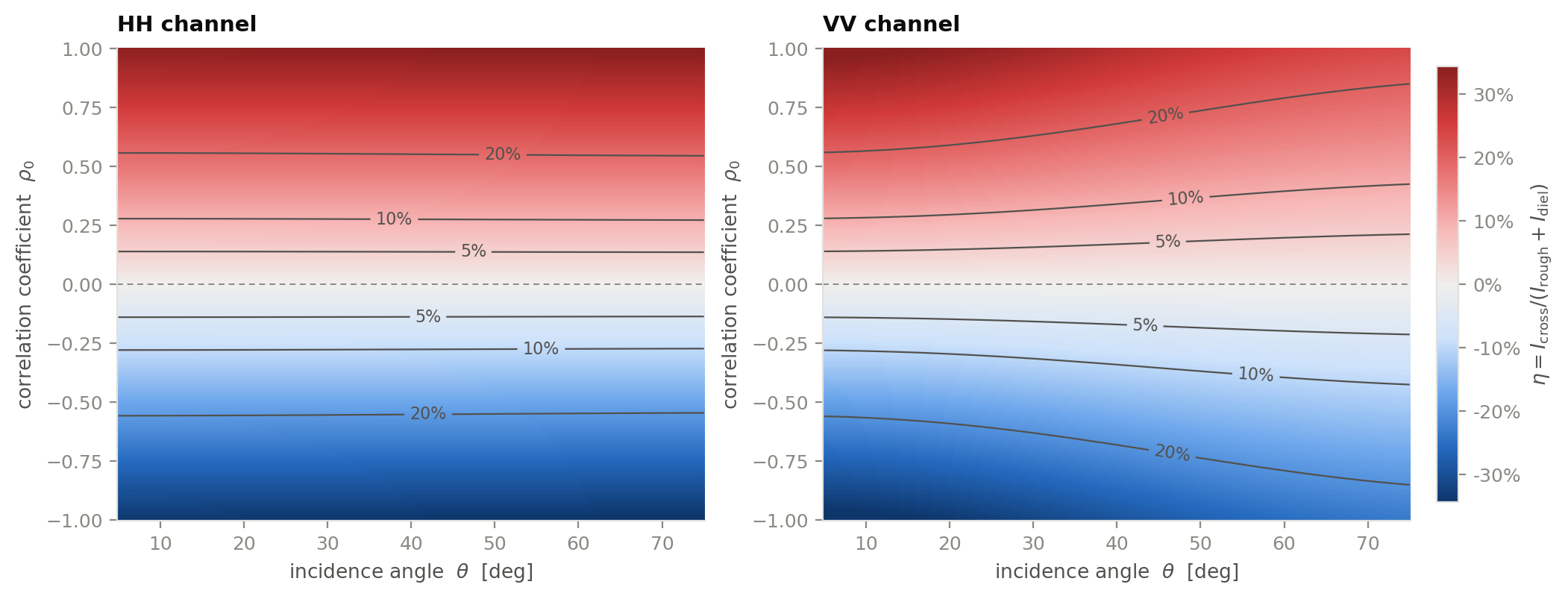}
    \caption{Fractional departure from additivity $\eta$, given by~(\ref{eta_def}), as a function of incidence angle and of the surface--dielectric correlation coefficient $\rho_0$. Contours at $5\%$, $10\%$ and $20\%$. Red is enhancement, blue suppression.}
    \label{fig:eta}
\end{figure}

If $\eta=0$, we are treating with exact additivity, while $\eta>0$ gives an enhancement and $\eta<0$ a suppression of the backscattered power. In order to analyze this error, we present in Figure~\ref{fig:eta} $\eta(\theta,\rho_0)$ for both channels. This fractional error reaches a maximum of $36\%$ for both channels at $|\rho_0|=1$, and, for $|\rho_0|\gtrsim0.14$, the error exceeds $5\%$ of the power for an additive scattering model, which represents the order of radiometric accuracy of a calibrated SAR (0.21~dB).

Furthermore, in Figure \ref{fig:eta} two features are worth noting. First, $\eta$ changes sign with $\rho_0$: the coupling can either enhance or suppress the return, and an additive model cannot reproduce this even qualitatively, since it predicts zero in both cases. Second, the dependence on the angle of incidence is quite moderate: for the HH channel, it is essentially absent (according to the flatness seen in Figure~\ref{fig:composition}), while in the VV channel, $\eta$ decreases from $36\%$ at $5^\circ$ to $24\%$ at $75^\circ$, following the imbalance between the two mechanisms as the roughness is predominant.

The behavior of $\eta$ is straightforward to understand: since $I_{\rm cross}\propto\rho_0\,s\,s_\vep$ while $I_{\rm rough}\propto s^2$ and
$I_{\rm diel}\propto s^2_\vep$, the cross term is a geometric mean of the other two. It is therefore largest when the two mechanisms are comparable and is hidden whenever one of them dominates: an imbalance of two orders of magnitude between $I_{\rm rough}$ and $I_{\rm diel}$ costs roughly an order of magnitude in $\eta$, however large $\rho_0$ may be. 

\begin{figure}[htbp]
    \centering
    \includegraphics[width=\textwidth]{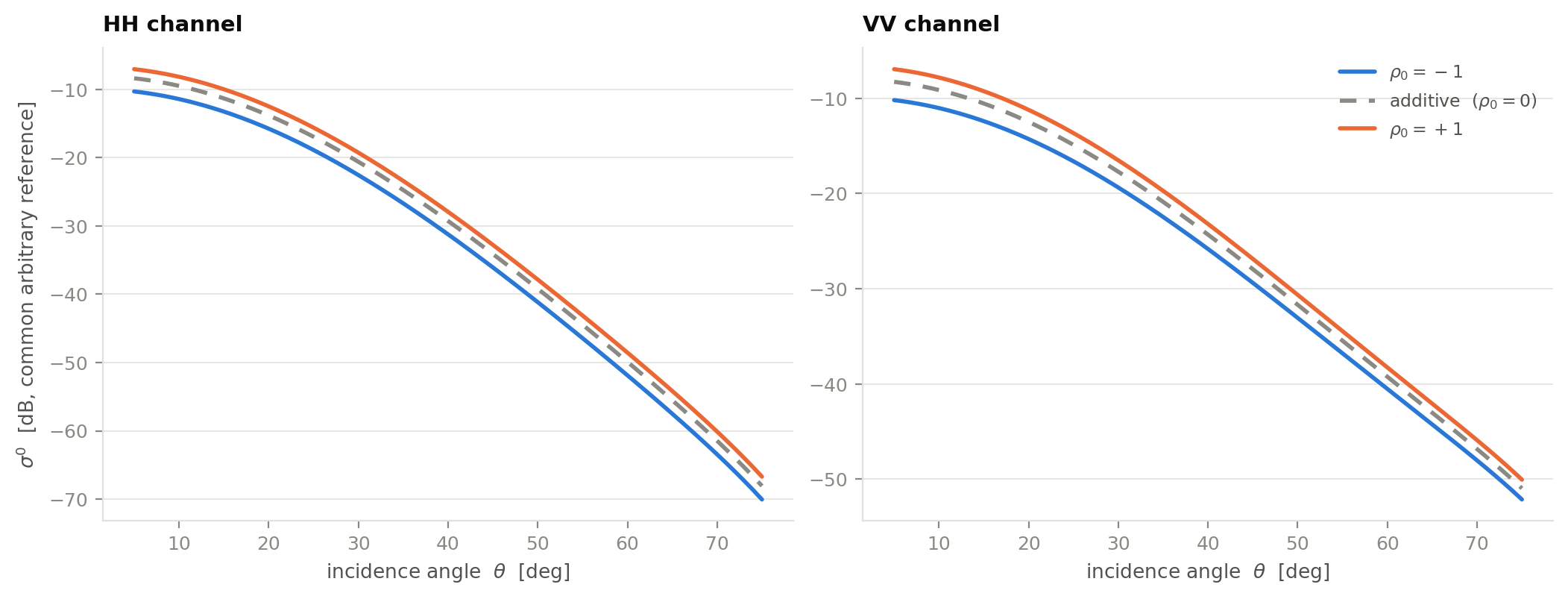}
    \caption{Backscattering cross-section $\sigma^0$, in dB relative to a common arbitrary reference, as a function of incidence angle, for the additive model ($\rho_0=0$, dashed) and for maximal enhancement/suppression ($\rho_0=\pm1$, solid). The vertical offset between the three curves is
    $10\log_{10}(1+\eta)$, the same quantity mapped in Figure~\ref{fig:eta}.}
    \label{fig:sigma0_curves}
\end{figure}

Finally, in Figure~\ref{fig:sigma0_curves} we show the same effect directly in the radar cross-section itself rather than in the dimensionless ratio $\eta$: the backscattering curves for $\rho_0=\pm1$ bracket the additive prediction by a nearly constant offset across the whole angular range, in both channels, confirming that the coupling rescales the level of $\sigma^0(\theta)$ without distorting its shape.

\subsection{The coherence condition on the correlated layer}

\begin{figure}[htbp]
    \centering
    \includegraphics[width=\textwidth]{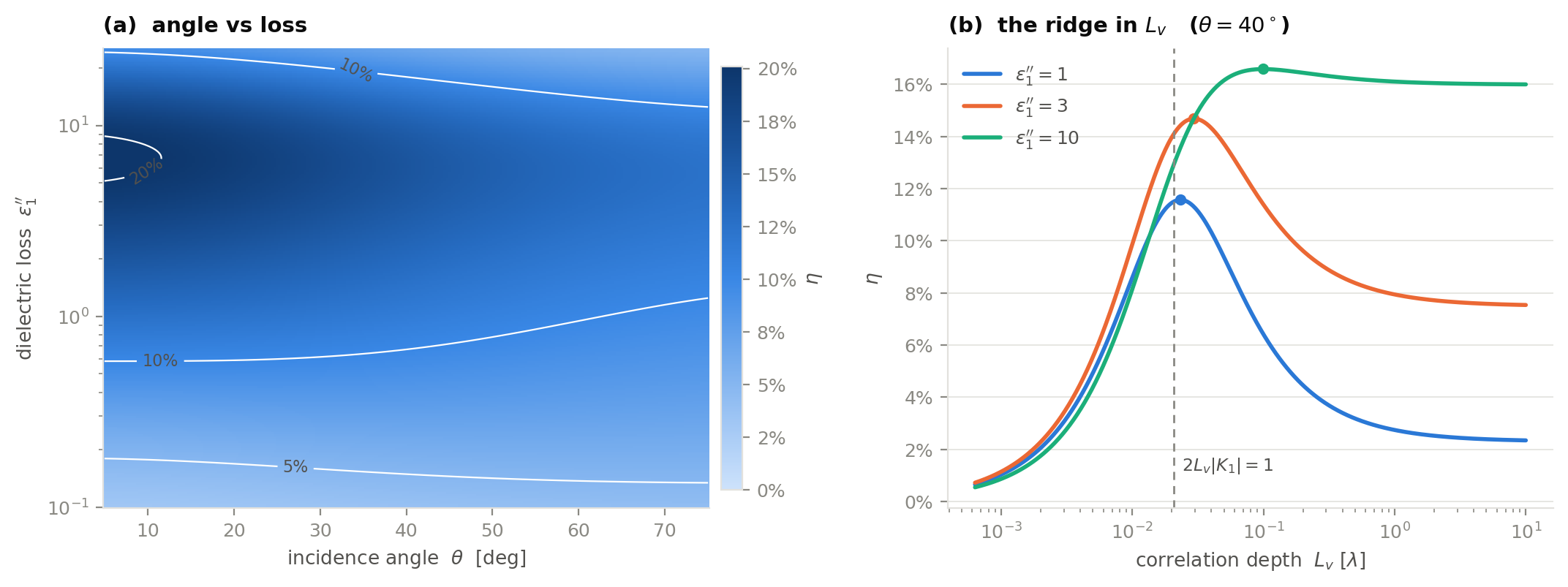}
    \caption{(a) $\eta$ in the VV channel as a function of incidence angle and dielectric loss, at $\rho_0=0.5$ and $\epsilon'_1=15$. (b) $\eta$ at
    $\theta=40^\circ$ as a function of the correlation depth $L_v$, for three values of the loss; dots mark the maxima and the dashed line is the coherence scale $2L_v|K_1|=1$. The crest is a property of the dielectric medium.}
    \label{fig:conditions}
\end{figure}

The second condition is specific to the three-dimensional medium model and cannot arise in a columnar one. The vertical structure of the cross-correlation enters in (\ref{HH_O1}) and (\ref{VV_O1}) only through

\begin{equation}
    \Omega_{h\vep}\;\propto\;\frac{L_v}{1+2L_v\left(K''_1+\imath K'_1\right)}\,,
    \label{Omega_hvep}
\end{equation}

which vanishes as $L_v\to0$ and saturates as $L_v\to\infty$. The physics of the two limits is clear: a correlated layer much thinner than the wavelength in the medium confines a small volume to contribute, while across a layer much thicker than the wavelength the phase $2L_vK'_1$ presents in (\ref{HH_meansquare}) and (\ref{VV_meansquare}) rotates and the deep contributions interfere destructively. Between the two there is a crest located at $2L_v\left\vert K_1\right\vert\sim1$, as shown in Figure~\ref{fig:conditions}(b). Numerically, it is equivalent to $2L_v|K_1|=1.1$--$1.5$ for $\epsilon''_1/\epsilon'_1\lesssim0.2$. The position of this crest is fixed by the dielectric medium and is almost independent of the incidence angle: for $\epsilon_1=15+3\imath$ it moves less than $4\%$ between $20^\circ$ and $60^\circ$. Its prominence, on the other hand, depends strongly on the losses in the dielectric: for $\epsilon''_1=1$ the peak is a factor of $5$ above the large-$L_v$ plateau, for $\epsilon''_1=3$ a factor of $2$, and by $\epsilon''_1=10$ vanishes: a wave that is absorbed within a fraction of a wavelength never penetrates deep enough for the phase to rotate.

Figure~\ref{fig:conditions}(a) shows the complementary dependence on the dielectric loss at fixed $L_v$. Increasing $\epsilon''_1$ enhances $\eta$ over more than a decade, up to $\epsilon''_1\approx10$. The reason is that the volume term integrates the whole illuminated depth, $I_{\rm diel}$ falling off at least as fast as $1/K''_1$, whereas the cross term only weighs the layer $|\z|\lesssim L_v$ adjacent to the interface: as the losses
grow, the volume contribution is extinguished first, and the near-surface coupling is left exposed.

\subsection{Consequence for retrieval: a bias in \texorpdfstring{$s_\vep$}{s\_eps}} \label{sec:bias}

As defined by (\ref{eta_def}), the quantity $\eta$ is not itself observable, since it presupposes the knowledge of the three contributions separately. However, if synthetic data generated by a correlated target are inverted with an additive model (i.e. $\rho_0=0$), then an actual error is introduced. If the amplitude of the dielectric fluctuations is fitted under this assumption, the cross term is absorbed into $s_\vep$, producing an effective $s^{\rm eff}_{\vep}$ amplitude for the  dielectric fluctuations. Due to $I_{\rm diel}\propto s^2_\vep$, an effective amplitude for the additive scattering model can be arranged as

\begin{equation}
    \frac{s^{\rm eff}_\vep}{s_\vep}=\sqrt{1+\frac{I_{\rm cross}}{I_{\rm diel}}}\,.
    \label{bias_def}
\end{equation}

\begin{figure}[htbp]
    \centering
    \includegraphics[width=\textwidth]{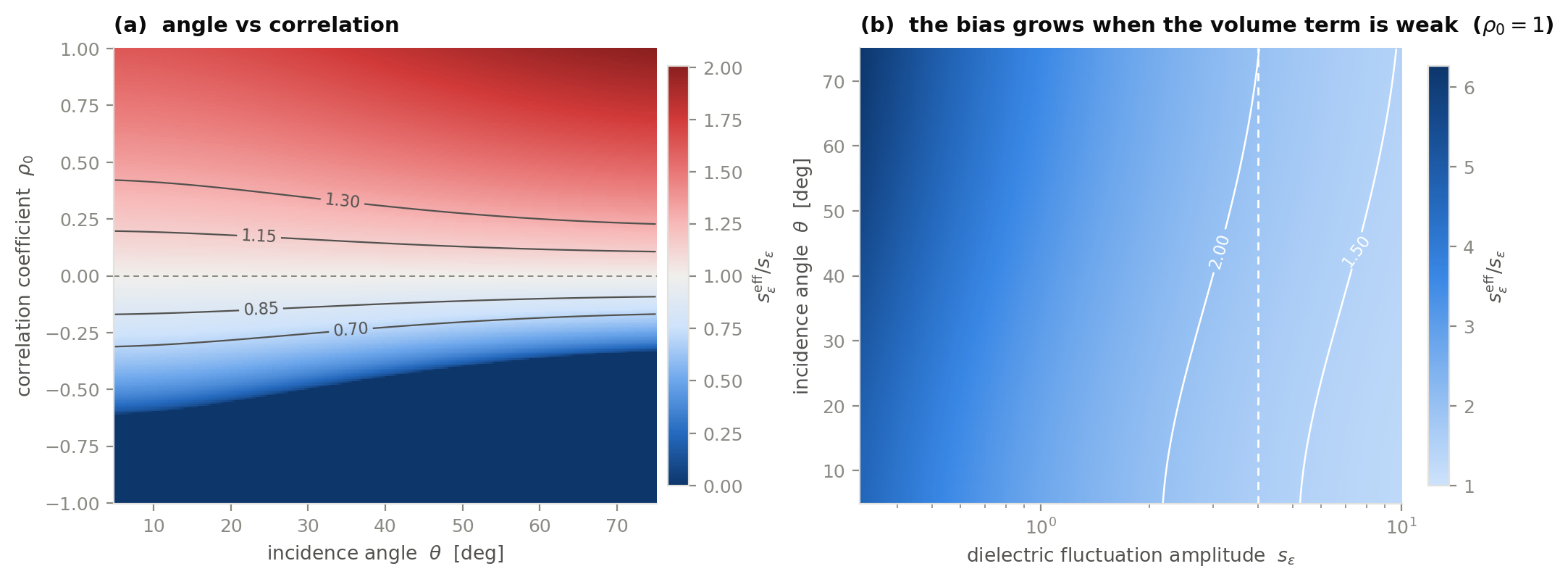}
    \caption{Bias factor $s^{\rm eff}_\vep/s_\vep$, eq.~(\ref{bias_def}), incurred when data from a correlated target are inverted with an additive model.
    (a) Against incidence angle and correlation coefficient. (b) Against the true fluctuation amplitude $s_\vep$ at $\rho_0=1$; the dashed line marks the value used in the other figures. The bias is worst where the volume term is weakest.}
    \label{fig:bias}
\end{figure}

The denominator here is $I_{\rm diel}$ alone, not $I_{\rm rough}+I_{\rm diel}$, so the bias is substantially greater than $\eta$ and grows when the dielectric term is weak. In Figure~\ref{fig:bias} we map this effect just for the VV channel. In the parameters used throughout, the retrieved amplitude is in error by $62$--$101\%$ at $\rho_0=1$; reducing $s_\vep$ to unity, so that the volume term is the weakest of the two, the retrieved amplitude exceeds the true one by a factor of $2.8$--$3.6$, and for $s_\vep=0.5$ by a factor of $3.8$--$5.0$. A negative $\rho_0$ biases the retrieval in the other way: at these parameters $\left\vert I_{\rm cross}\right\vert$ exceeds $I_{\rm diel}$, so that for $\rho_0\lesssim-0.33$ at some angles (and at every angle once $\rho_0\lesssim-0.61$) $1+\frac{I_{\rm cross}}{I_{\rm diel}}<0$ and therefore the additive model has no solution: an anti-correlated target returns less power than its roughness alone would, and a sum of two non-negative contributions cannot reproduce that for any value of $s_\vep$.

To close this section, we make two remarks. First, the bias is a systematic error, not a loss of precision: it does not average down over repeated acquisitions, and it cannot be diagnosed from the angular shape of $\sigma^0$, for the reason given in Figure~\ref{fig:composition}. Second, since the coupling does not have an angular signature of its own and does not appear in the polarization ratio, exposing it requires an independent constraint on $s_\vep$ or a second frequency for which the condition $2L_v\left\vert K_1\right\vert\sim1$ changes the value of $L_V/\lambda$ where the crest occurs. 
The frequency dependence of the coherence condition is, in that sense, the most promising handle that the present result offers.

\section{Conclusions} \label{sec:conclusions}

We have derived the first-order incoherent backscattering cross-section of a random rough surface bounding a half-space of randomly fluctuating permittivity in a formulation in which the two mechanisms and their interference arise from a single calculation.

The non-orthogonal transformation $z=\z+h(\vx)f(\z)$ maps the rough boundary onto a flat interface and converts the geometric roughness into volume sources of the same Helmholtz equation, and in the same region, as the dielectric fluctuations. Then, both kinds of source are propagated by the same Green's function. As a consequence, the interference between roughness and permittivity emerges as a term of the mean square rather than as an imposed addition.

The proposed method was verified against the standard benchmark for the scattering of random rough surfaces. The geometric part of the scattering amplitudes reproduces the Small Perturbation Method kernel and, what matters to the method, it does so for an arbitrary profile function $f(\z)$. Regarding the profile used in the change of coordinates, it is only required that the function $f(\z)$ vanishes far away from the true rough interface and also that it be continuous along it. No further conditions on $f(\z)$ are needed.

Beyond the benchmark, the results presented in Section~\ref{sec:results} can be summarized mainly in three points: (i) the fractional departure from additivity $\eta$ reaches $36\%$ in both co-polarized channels at $|\rho_0|=1$, and exceeds $5\%$ (the order of the radiometric accuracy of a calibrated SAR) for $|\rho_0|\gtrsim0.14$. Depending on the sign of the correlation, it either enhances or suppresses the return, which an additive model cannot reproduce even qualitatively; (ii) since $I_{\rm cross}\propto\rho_0 s\,s_\vep$ is a geometric mean of $I_{\rm rough}\propto s^2$ and $I_{\rm diel}\propto s^2_\vep$, the coupling is buried whenever either mechanism dominates the other. Furthermore, the cross term is governed by 
$\Omega_{h\vep}\propto L_v/[1+2L_v(K''_1+\imath K'_1)]$, which vanishes for a correlated layer too thin to enclose any volume and falls back to a plateau once the layer is thick enough for the phase to rotate across it, with a peak between at $2L_v|K_1|\sim1$. This peak is a property of the random dielectric medium, rather than of the observation geometry, and it is sharp only in a low-loss medium. A condition of this kind cannot appear in a columnar medium model, which has no vertical scale to vary; (iii) because the coupling rescales the level rather than distorting the shape of the backscattering cross section, it is absorbed by the fitted parameters. If the amplitude of the dielectric fluctuations is retrieved under the assumption $\rho_0=0$, the whole cross term is loaded onto $s_\vep$, overestimating the true amplitude of the dielectric fluctuations, or, for a sufficiently negative $\rho_0$, the additive model does not merely misestimate $s_\vep$: it has no solution at all, the target returning less power than its own roughness would.

Two limitations should be clearly stated. The results are first order in both random fields, so they do not describe the multiple-scattering interaction between surface and volume that arises when the fluctuations are not small, a mechanism which is present even for statistically independent randomness \cite{mudaliar1995wave}, and which is distinct from the statistical coupling computed here. Moreover, the correlation functions have been taken as Gaussian in the lateral coordinates with an exponential vertical profile; the coherence condition, $2L_v\left\vert K_1\right\vert\sim1$, depends on the existence of a vertical scale and not on its particular shape, but the numerical factor $2L_v|K_1|\simeq1.1$--$1.5$ does. The construction itself is not tied to the first order: the sources of Section~\ref{sec:first_order_sources} are defined at arbitrary order, and the extension is a matter of carrying the expansion further rather than of reformulating the problem.

\section*{Acknowledgments}
\noindent
M. F. acknowledges financial support from Consejo Nacional de Investigaciones Cient\'ificas y T\'ecnicas (CONICET) Grant No. PIP 11220220100465. \\
E. C. acknowledges financial support from Universidad de Buenos Aires through Grant No. UBACYT 20020170100129BA and CONICET Grant No. PIP2017/19:11220170100817CO.

\bigskip

\section*{Note on the use of AI assistance} 

The explicit evaluation of the integrals collected in this appendix, and the comparison of the resulting geometric amplitudes with the corresponding kernels of the Small Perturbation Method were carried out with the assistance of a large language model (Claude, Anthropic). Every expression obtained in this way was checked independently: symbolically with \texttt{sympy} and numerically with \texttt{mpmath}, over a range of incidence angles, dielectric permittivity and profile functions $f(\z)$. The verification scripts, together with the symbolic derivation of the first-order sources of Section~\ref{sec:first_order_sources}, are provided in the accompanying code repository. The authors have checked every result reported here and are solely responsible for its content.

\bigskip

\section*{Data availability statement}
No measured data were used in this study. The symbolic derivation of the first-order
sources of Section~\ref{sec:first_order_sources}, the verification of the geometric amplitudes (\ref{IHH_exact})
and (\ref{eq:IVV_exact}), and the scripts that generate every figure in this
paper are openly available at
\url{https://github.com/mfranco82/Non-additive-surface-and-volume-backscattering},
archived at version v1.3.0, \texttt{doi:10.5281/zenodo.22967900} \cite{franco2026code}.

\section*{Disclosure statement}
No potential conflict of interest was reported by the authors.
%

\section*{ORCID}
\noindent
Mariano Franco \orcidicon{0000-0001-7611-1688} \href{https://orcid.org/0000-0001-7611-1688}{https://orcid.org/0000-0001-7611-1688} \\
Esteban Calzetta \orcidicon{0000-0002-3083-3420} \href{https://orcid.org/0000-0002-3083-3420}{https://orcid.org/0000-0002-3083-3420}

\appendix 
\section{The two correlation integrals} \label{appendix_inegrals}

The two geometric amplitudes, $I^h_{HH}$ of (\ref{IHH_exact}) and $I^h_{VV}$ of (\ref{eq:IVV_exact}), are integrals over the profile $f(\z)$ of products of zeroth-order fields. Their evaluation is elementary but long: each one splits into the two half-spaces, each half-space produces a moment 

\begin{align}
        \mathcal{F}_>(\lambda)&=\int^\infty_0d\z\,f(\z)e^{\imath\lambda\z} \nonumber \\
        \mathcal{F}_<(\lambda)&=\int^0_{-\infty}d\z\,f(\z)e^{\imath\lambda\z} \nonumber
\end{align}

evaluated at $\lambda=\pm2K_0$ above and $\lambda=-2K_1$ below, and these moments cancel --- in HH within a single integral, in VV between $I_\parallel$ and $I_\perp$, coefficient by coefficient and not merely in total. We do not reproduce that algebra here. It is carried out in full, twice and independently --- symbolically in $K_0,K_1,k,\epsilon_1$ and numerically by quadrature with an explicit profile having $f'(0)\neq0$ --- in the notebook \texttt{verify\_appendix.ipynb} of the accompanying repository \cite{franco2026code}, which also verifies there the three statements the text makes about it: that the moments cancel coefficient by coefficient, that the value of $f'(0)$ is immaterial, and that the resulting amplitudes close on the SPM kernels identically in angle and permittivity.

What that calculation cannot supply, because it concerns the statistics of the medium rather than the coordinate map, are the two $\z$-integrals produced by the correlation functions (\ref{diel_corr}) and (\ref{h_vep_corr}) when the mean squares are formed. They carry the whole dependence of the result on the vertical structure of the medium, they are the same in both polarizations --- the channels differ only by the prefactor multiplying them --- and they are evaluated once here.

\subsection{The dielectric moment} \label{app:eps_integral}

The second line of (\ref{VV_meansquare}), and its counterpart in the HH channel, require

\begin{equation}
    J_{\vep} \;\equiv\; \iint^0_{-\infty} d\z\,d\z'\;e^{-|\z-\z'|/l_v}\;
    e^{-2\imath\left(K_1\z-K^*_1\z'\right)}\,,
    \label{eq:Jeps_def}
\end{equation}

the exponential being the vertical part of (\ref{diel_corr}) and the phase that of the two propagators. Write $K_1=K'_1+\imath K''_1$ with $K''_1>0$ and change variables to $u=\z-\z'$, $v=\z+\z'$, for which $d\z\,d\z'=\tfrac12\,du\,dv$ and the phase becomes $-2\imath K'_1u+2K''_1v$. The domain $\z,\z'<0$ is $v<-|u|$, so the $v$-integral is elementary and leaves an absorption factor $e^{-2K''_1|u|}$ behind,

\begin{align}
    J_{\vep} &= \frac{1}{2}\int^{\infty}_{-\infty}du\;e^{-|u|/l_v}\,e^{-2\imath K'_1u}
    \int^{-|u|}_{-\infty}dv\;e^{2K''_1v}
    \;=\;\frac{1}{4K''_1}\int^{\infty}_{-\infty}du\;
    e^{-|u|\left(1/l_v+2K''_1\right)}\,e^{-2\imath K'_1u} \nn
    &= \frac{1}{2K''_1}\;\frac{\tilde{l}_v}{1+\left(2K'_1\tilde{l}_v\right)^2}\,,
    \qquad
    \tilde{l}_v\;\equiv\;\frac{l_v}{1+2K''_1\,l_v}\,.
    \label{eq:Jeps}
\end{align}

Two scales are therefore at play, and they enter differently. The absorption length $1/(2K''_1)$ sets the overall depth of the layer that radiates, and appears as the prefactor. The vertical correlation length enters only through the effective length $\tilde{l}_v$, which is $l_v$ itself when the medium is transparent over one correlation length ($2K''_1l_v\ll1$) and saturates at $1/(2K''_1)$ otherwise, and through the Lorentzian $1+(2K'_1\tilde{l}_v)^2$, which is the Bragg factor of the vertical structure: the fluctuations radiate coherently only while $2K'_1\tilde{l}_v\lesssim1$, and are
averaged out beyond that. In the lossless limit $K''_1\rightarrow0$ the integral diverges, as it must: an unbounded non-absorbing half-space of fluctuations radiates without bound.

Multiplying (\ref{eq:Jeps}) by $k^4_i|T_H|^4/(2K_0)$ and by the lateral part $(s^2_{\vep}l^2_r/4\pi)e^{-(kl_r)^2}$ of (\ref{diel_corr}) gives the second line of (\ref{HH_O1}); multiplying it instead by $|\mathcal{N}_0|^2$ of (\ref{Nkernel}) gives the second line of (\ref{VV_O1}).

\subsection{The roughness--dielectric moment} \label{app:heps_integral}

The third line of (\ref{VV_meansquare}), and its counterpart in HH, require the single integral

\begin{align}
    J_{h\vep} & \equiv\; \int^0_{-\infty} d\z\;e^{-|\z|/L_v}\,e^{2\imath K^*_1\z}
    \;=\; \int^0_{-\infty} d\z\;e^{\z/L_v+2\left(K''_1+\imath K'_1\right)\z}
    \nn 
    J_{h\vep} &= \frac{L_v}{1+2L_v\left(K''_1+\imath K'_1\right)}\,,
    \label{eq:Jheps}
\end{align}

with $L_v$ the vertical decay length of (\ref{h_vep_corr}). Only one propagator appears here, and only one integration: the roughness lives at the interface, so the correlation $W_{h\vep}$ ties the dielectric fluctuation at depth $\z$ to a surface quantity, not to a second fluctuation at another depth. The integral converges at $\z\rightarrow-\infty$ for $K''_1\geq0$ even for $L_v\rightarrow\infty$, in contrast with (\ref{eq:Jeps}): a correlation that does not decay with depth still gives a finite cross term, damped by absorption alone.

Unlike $J_{\vep}$, $J_{h\vep}$ is complex, and the condition for its phase not to have averaged the cross term away is $2L_v|K_1|\lesssim1$ --- the coherence condition discussed in Section~\ref{sec:results}. Multiplying (\ref{eq:Jheps}) by $-2\,k^2_i\,I^h_{HH}(-\vk)\,(T^*_H)^2/\sqrt{2K_0}$ and by
$\rho_0(s_{\vep}sL^2_r/4\pi)e^{-(kL_r)^2}$, and taking twice the real part, gives the third line of (\ref{HH_O1}); the same operation with
$I^h_{VV}(-\vk)\,\epsilon^*_1k^2_ik^2T^*_V/(K^*_1)^2$ gives the third line of (\ref{VV_O1}).

\bibliographystyle{ieeetr}
\bibliography{biblio}

\end{document}